%% file: paper.tex
\pdfoutput=1
\newcommand*{\EXTENDEDVERSION}{}

\ifdefined\EXTENDEDVERSION
\documentclass[sigconf,svgnames,table,anonymous=false]{acmart}
\else
\documentclass[sigconf,svgnames,table,review,anonymous]{acmart}
\fi

\setcopyright{rightsretained}

\ifdefined\EXTENDEDVERSION
\renewcommand\footnotetextcopyrightpermission[1]{}
\fi

\ifdefined\EXTENDEDVERSION
\else
\acmConference[Anonymous Submission to ACM CCS 2018]{ACM Conference on Computer and Communications Security}{15-19 October 2018}{Toronto, Canada}
\acmYear{2018}
\acmISBN{} 
\acmDOI{} 
\fi

\startPage{1}

\setcopyright{none}
\usepackage{booktabs}   
\usepackage{subcaption} 
\usepackage{xcolor}
\usepackage{textcomp}
\usepackage{pgfpages}
\usepackage{pgfplots}
\usepackage{tikz}
\usepackage{enumerate}
\usepackage{algorithmicx}
\usepackage{algorithm}
\usepackage[noend]{algpseudocode}
\usepackage{multirow}
\usepackage{amsmath}
\usepackage{listings}

\definecolor{light-gray}{gray}{0.77}
\definecolor{BrickRed}{rgb}{0.8, 0.25, 0.33}

\definecolor{lightorange}{HTML}{FFB74D}
\makeatletter
\newenvironment{btHighlight}[1][]
{\begingroup\tikzset{bt@Highlight@par/.style={#1}}\begin{lrbox}{\@tempboxa}}
{\end{lrbox}\bt@HL@box[bt@Highlight@par]{\@tempboxa}\endgroup}

\newcommand\btHL[1][]{%
  \begin{btHighlight}[#1]\bgroup\aftergroup\bt@HL@endenv%
}
\def\bt@HL@endenv{%
  \end{btHighlight}%
  \egroup
}
\newcommand{\bt@HL@box}[2][]{%
  \tikz[#1]{%
    \pgfpathrectangle{\pgfpoint{0.3pt}{0pt}}{\pgfpoint{\wd #2}{\ht #2}}%
    \pgfusepath{use as bounding box}%
    \node[anchor=base west,fill=lightorange,outer sep=0pt,inner xsep=0.3pt,inner ysep=0pt,minimum height=\ht\strutbox+0.3pt,#1]{\raisebox{0.3pt}{\strut}\strut\usebox{#2}};
  }%
}
\makeatother

\usetikzlibrary{calc,trees,positioning,arrows,chains,shapes.geometric,%
  decorations.pathreplacing,decorations.pathmorphing,decorations.text,shapes,%
  matrix,shapes.symbols,patterns,shadows}

\newenvironment{customlegend}[1][]{%
    \begingroup
    \csname pgfplots@init@cleared@structures\endcsname
    \pgfplotsset{#1}%
}{%
    \csname pgfplots@createlegend\endcsname
    \endgroup
}

\def\addlegendimage{\csname pgfplots@addlegendimage\endcsname}

\pgfplotsset{compat=newest}

\newcommand*\Let[2]{\State #1 $\gets$ #2}
\newcommand*\LetHL[2]{\State {\btHL[fill=light-gray] #1 $\gets$ #2}}
\newcommand*\Fcall[1]{\textsc{#1}}
\newcommand*\IfHL[1]{\State {\btHL[fill=light-gray] \textbf{if} #1 \textbf{then}}}
\newcommand*\ElseHL{\State {\btHL[fill=light-gray] \textbf{else}}}
\algrenewcommand\alglinenumber[1]{\tiny\color{Black!70}{#1}}
\algrenewcommand\algorithmicforall[2]{\textbf{for} $i=$ #1 \textbf{to} #2}
\algrenewtext{ForAll}{\algorithmicforall}
\algnewcommand\algorithmicswitch{\textbf{switch}}
\algnewcommand\algorithmiccase{\textbf{case}}
\algdef{SE}[SWITCH]{Switch}{EndSwitch}[1]{\algorithmicswitch\ #1\ \algorithmicdo}{\algorithmicend\ \algorithmicswitch}%
\algdef{SE}[CASE]{Case}{EndCase}[1]{\algorithmiccase\ #1}{\algorithmicend\ \algorithmiccase}%
\algtext*{EndSwitch}%
\algtext*{EndCase}%
\algdef{SE}[SUBALG]{Indent}{EndIndent}{}{\algorithmicend\ }%
\algtext*{Indent}
\algtext*{EndIndent}

\input{solidity-highlighting.tex}

\lstdefinestyle{basic}{%
  morekeywords     = [1]{var},%
  morekeywords     = [2]{},%
  keywordstyle     = [2]\color{teal}\bfseries,%
  morekeywords     = [3]{minimize},%
  keywordstyle     = [3]\color{BrickRed}\bfseries,%
  keywordstyle     = \bfseries\color{DarkBlue},%
  commentstyle     = \small\ttfamily\color{Black!70},%
  basicstyle       = \small\ttfamily,%
  emph             = {int,char,double,float,unsigned,void,bool},%
  emphstyle        = {\color{teal}\bfseries},%
  columns          = [c]fixed,%
  aboveskip        = 0mm,%
  belowskip        = 2mm,%
  keepspaces       = true,%
  mathescape       = true,%
  escapechar       = ¤,%
  tabsize          = 2,%
  numbers          = left,%
  numberstyle      = \tiny\color{Black!70},%
  numbersep        = 4pt,%
  stepnumber       = 1,%
  firstnumber      = 1,%
  showstringspaces = false,%
  captionpos       = b,%
  extendedchars    = true,%
  upquote          = true,%
  abovecaptionskip = 0mm,%
  belowcaptionskip = 0mm,%
  moredelim        = **[is][{\btHL[fill=light-gray]}]{°}{°},%
}

\lstdefinestyle{clang}{%
  language         = C,%
  style            = basic,%
}

\lstdefinestyle{sol}{%
  language         = Solidity,%
  style            = basic,%
}

\newcommand\code[1]{\lstinline[style=clang]{#1}}

\begin{document}

\title[Learning Inputs in Greybox Fuzzing]{Learning Inputs in Greybox Fuzzing}


\author{Valentin W{\"u}stholz}
\affiliation{}
\email{wuestholz@gmail.com}

\author{Maria Christakis}
\affiliation{
  \institution{MPI-SWS, Germany}
}
\email{maria@mpi-sws.org}

\begin{abstract}
Greybox fuzzing is a lightweight testing approach that effectively
detects bugs and security vulnerabilities. However, greybox fuzzers
randomly mutate program inputs to exercise new paths; this makes it
challenging to cover code that is guarded by complex checks.

In this paper, we present a technique that extends greybox fuzzing
with a method for learning new inputs based on already explored
program executions. These inputs can be learned such that they guide
exploration toward specific executions, for instance, ones that
increase path coverage or reveal vulnerabilities. We have evaluated
our technique and compared it to traditional greybox fuzzing on 26
real-world benchmarks. In comparison, our technique significantly
increases path coverage (by up to 3X) and detects more bugs (up to
38\% more), often orders-of-magnitude faster.
\end{abstract}




\maketitle

\section{Introduction}

There is no doubt that test case generation helps find bugs and
security vulnerabilities, and therefore, improve software quality. It
is also clear that manual test case generation is a cumbersome and
very time-consuming task. As a consequence, there has emerged a wide
variety of automatic test case generation tools, ranging from random
testing~\cite{ClaessenHughes2000,CsallnerSmaragdakis2004,PachecoLahiri2007},
over greybox fuzzing~\cite{AFL,LibFuzzer}, to dynamic symbolic
execution~\cite{GodefroidKlarlund2005,CadarEngler2005}.

Random
testing~\cite{ClaessenHughes2000,CsallnerSmaragdakis2004,PachecoLahiri2007}
and blackbox fuzzing~\cite{PeachFuzzer,ZzufFuzzer} generate random
inputs to a program, run the program with these inputs, and check
whether any bugs were found, e.g., whether the program crashed or a
user-provided property was violated. Despite their practicality, the
effectiveness of random testing and blackbox fuzzing, that is, their
ability to explore new paths, is limited. The search space of valid
program inputs is typically huge, and a random exploration can only
exercise a small fraction of (mostly shallow) program paths.

At the other end of the spectrum, dynamic symbolic
execution~\cite{GodefroidKlarlund2005,CadarEngler2005} and whitebox
fuzzing~\cite{GodefroidLevin2008,CadarDunbar2008,GaneshLeek2009}
repeatedly run a program, both concretely and symbolically. At
runtime, they collect symbolic constraints on program inputs from
branch statements along the execution path. These constraints are then
appropriately modified and a constraint solver is used to generate new
inputs, thereby steering execution toward another program
path. Although these techniques are very effective in covering new
paths, they simply cannot be as efficient and scalable as other test
case generation techniques that do not spend any time on program
analysis and constraint solving.

Greybox fuzzing~\cite{AFL,LibFuzzer} lies in the middle of the
spectrum between performance and effectiveness in discovering new
paths. It does not require program analysis or constraint solving, but
it relies on a lightweight program instrumentation that allows the
fuzzer to tell when an input exercises a new path. In other words, the
instrumentation is useful in computing a unique identifier for each
explored path in a program under test. American Fuzzy Loop
(AFL)~\cite{AFL} is a prominent example of a state-of-the-art greybox
fuzzer that has detected numerous bugs and security
vulnerabilities~\cite{AFL-Bugs}.

A greybox fuzzer, like AFL, starts by running the program under test
with a seed input. During execution of the program, the fuzzer
computes the identifier of the path that is currently being
explored. It then mutates the seed slightly to generate a new input
and runs the program again. If a new path is exercised, the input is
added to a set of seeds, otherwise it is discarded. From this set of
seeds, the fuzzer picks new inputs to mutate. The process is repeated
until a timeout is reached or until the user aborts the exploration.

Despite the fact that greybox fuzzing strikes a good balance between
performance and effectiveness in automatic test case generation, the
inputs are still mutated in a random way, for instance, by flipping
random bits. As a result, many generated inputs exercise the same
program paths. To address this problem, there have emerged techniques
that direct greybox fuzzing toward low-frequency
paths~\cite{BoehmePham2016}, vulnerable paths~\cite{RawatJain2017},
deep paths~\cite{SparksEmbleton2007}, or specific sets of program
locations~\cite{BoehmePham2017}.
Specifically, such techniques have focused on which seed inputs to
prioritize and which parts of these inputs to mutate.

\vspace{1em}
\paragraph{\textbf{Our approach}} In this paper, we present an
orthogonal approach that systematically learns new inputs for the
program under test with the goal of increasing the performance and
effectiveness of greybox fuzzing. In contrast to existing work, our
approach suggests concrete input values based on information from
previous executions, instead of performing arbitrary mutations. The
new inputs are learned in a way that aims to direct greybox fuzzing
toward \emph{optimal} executions, for instance, executions that flip a
branch condition and thus increase path coverage. Our technique is
parametric in what constitutes an optimal execution, and in
particular, in what properties such an execution needs to satisfy.

More specifically, each program execution can be associated with zero
or more \emph{cost metrics}. A cost metric captures how close the
execution is to satisfying a given property at a given program
location.
Executions that minimize a cost metric are considered optimal with
respect to that metric.
For example, one could define a cost metric at each arithmetic
operation in the program such that the metric becomes minimal (i.e.,
equal to zero) when an execution triggers an arithmetic overflow.
Then, any execution that leads to an overflow is optimal with respect
to the cost metric it minimizes.
Note that cost metrics can either be computed automatically, as in the
example above, or provided by the user.

Our algorithm uses the costs that are computed along executions of the
program under test to learn new inputs that lead to optimal
executions. As an example, consider that method \code{foo}, shown
below, is the program under test. (The grey box on line~\ref{line:min}
should be ignored for now.)

\newpage

\vspace{0.7em}
\begin{lstlisting}[style=clang, xleftmargin=1.5em]
void foo(int x) {
  x = x - 7;
  °minimize(|x - 42|);° ¤\label{line:min}¤
  if x == 42 { vulnerability(); }
}
\end{lstlisting}
\noindent
The problem revealed by this code is typical of black- or greybox
fuzzers: it is difficult to generate a value for input \code{x} that
drives the program through the vulnerable path. For the sake of this
example, we assume the cost metric to be the absolute value of \code{x
  - 42}, shown on line~\ref{line:min}, which defines optimal
executions as those that cover the then-branch of the if-statement.
Now, let us imagine that none of the first few executions of this
program cover the then-branch. Although these executions are not
optimal with respect to the cost metric of line~\ref{line:min}, they
are still useful in learning a correlation between the input values of
\code{x} and their corresponding costs (as computed on
line~\ref{line:min}). Based on this approximate correlation, our
algorithm proposes a new value for \code{x} (namely 49), which
minimizes the cost on line~\ref{line:min}. In contrast, existing
greybox fuzzing techniques randomly mutate \code{x} hoping to cover
the vulnerable path.

Another way to think about our algorithm is the following. Given a set
of cost metrics, the program under test is essentially a function from
$\mathit{inputs}$ to $\mathit{costs}$, $\mathit{costs} =
f(\mathit{inputs})$, which precisely describes the correlation between
inputs and the corresponding costs. In this context, our greybox
fuzzing algorithm collects data points of the form $(\mathit{inputs},
\mathit{costs})$ from executions of the program under test, and it
then approximates the function that correlates them. If the
approximation is precise enough, this technique is able to suggest
inputs that lead to optimal executions by purely reasoning about the
much simpler, approximated function.

We implemented our algorithm in a tool architecture that can be
instantiated with any cost metric. As a result, our architecture is
flexible enough to capture different notions of what constitutes an
optimal execution. In our instantiation of the tool architecture, we
consider two kinds of cost metrics, (1)~ones that are minimized when
execution flips a branch condition, and (2)~ones that are minimized
when execution is able to modify arbitrary memory locations. In
comparison to traditional greybox fuzzing, our approach is able to
significantly increase path coverage (by up to 3X) and detect many
more vulnerabilities (up to 38\% more), often orders-of-magnitude
faster.

\paragraph{\textbf{Contributions}} Our paper makes the following contributions:
\begin{itemize}
\item We introduce a greybox fuzzing algorithm that learns new inputs
  to guide exploration toward optimal program executions.

\item We implement this algorithm in a tool architecture that may be
  instantiated with different notions of optimality.

\item We evaluate our technique on 26 real-world benchmarks and
  demonstrate that it is more effective than traditional greybox
  fuzzing.
\end{itemize}

\paragraph{\textbf{Outline}} The next section reviews traditional greybox fuzzing and
gives an overview of our approach through a running example.
Sect.~\ref{sect:technique} explains the technical details of our
approach, while Sect.~\ref{sect:implementation} describes our
implementation. We present our experimental evaluation in
Sect.~\ref{sect:evaluation} and threats to the validity of our
experiments in Sect.~\ref{sect:threats}. We discuss related work in
Sect.~\ref{sect:relatedWork} and conclude in
Sect.~\ref{sect:conclusion}.

\section{Overview}
\label{sect:overview}

In this section, we first describe traditional greybox fuzzing and
then compare with our approach through a running example.

\subsection{Background: Greybox Fuzzing}
\label{subsect:fuzzing}

Fuzzing comes in three main flavors depending on how much the fuzzer
is allowed to look at the structure of the program under test.
\emph{Blackbox fuzzers}~\cite{PeachFuzzer,ZzufFuzzer}, like random
testing tools, execute the program under test without leveraging any
information about its structure. On the other hand, \emph{whitebox
  fuzzers}~\cite{GodefroidLevin2008,CadarDunbar2008,GaneshLeek2009},
like dynamic symbolic execution tools, use program analysis and
constraint solving to understand the entire structure of the
program under test and, therefore, explore more paths.

\emph{Greybox fuzzers}, like AFL~\cite{AFL} or
LibFuzzer~\cite{LibFuzzer}, leverage only some structure of the
program under test, which they obtain with lightweight program
instrumentation. By obtaining more information about the program
structure, greybox fuzzing can be more effective than blackbox fuzzing
in finding bugs. In comparison to whitebox fuzzing, greybox fuzzing
spends less time acquiring information about the program structure
and, as a result, it is more efficient.

\begin{algorithm}[t]
  \caption{Greybox Fuzzing}
  \label{alg:greyboxFuzzing}
  \hspace{-13em}\textbf{Input:} Program $\mathit{prog}$, Seeds $S$\\
  \begin{algorithmic}[1]
    \small
      \Let{$\mathit{PIDs}$}{\Fcall{RunSeeds}$(S, \mathit{prog})$}
      \While{$\neg$\Fcall{Interrupted}()}
      \Let{$\mathit{input}$}{\Fcall{PickInput}$(\mathit{PIDs})$}
        \Let{$\mathit{energy}$}{0}
        \Let{$\mathit{maxEnergy}$}{\Fcall{AssignEnergy}$(\mathit{input})$}
        \While{$\mathit{energy} < \mathit{maxEnergy}$}
          \Let{$\mathit{input'}$}{\Fcall{FuzzInput}$(\mathit{input})$}
          \Let{$\mathit{PID'}$}{\Fcall{Run}$(\mathit{input'}, \mathit{prog})$}
          \If{\Fcall{IsNew}($\mathit{PID'}, \mathit{PIDs}$)}
            \Let{$\mathit{PIDs}$}{\Fcall{Add}$(\mathit{PID'}, \mathit{input'}, \mathit{PIDs})$}
          \EndIf
          \Let{$\mathit{energy}$}{$\mathit{energy} + 1$}
        \EndWhile
      \EndWhile
  \end{algorithmic}
  \hspace{-12em}\textbf{Output:} Test suite \textsc{Inputs}$(\mathit{PIDs})$
\end{algorithm}

Alg.~\ref{alg:greyboxFuzzing} shows how greybox fuzzing works. The
fuzzer takes as input the program under test $\mathit{prog}$ and a set
of seeds $S$. It starts by running the program with the seeds, and
during each program execution, the instrumentation is able to capture
the path that is currently being explored and associate it with a
unique identifier $\mathit{PID}$ (line~1). Note that the
$\mathit{PIDs}$ data structure is a key-value store from a
$\mathit{PID}$ to an input that exercises the path associated with
$\mathit{PID}$.
Next, an input is selected for mutation (line~3); the selection can be
either random or based on heuristics, for instance, based on the
number of times an input has already been mutated. The selected input
is then assigned an ``energy'' value that denotes how many times the
selected input should be fuzzed (line~5).

The input is mutated (line~7), and the program is run with the new
input (line~8). If the program follows a path that has not been
previously explored, the new input is added to the test suite
(lines~9--10).
The above process is repeated until an exploration bound is reached
(line~2), for instance, a timeout, a maximum number of generated
inputs or seeds, etc.  The fuzzer returns a test suite containing one
test case for each program path that has been explored.
As shown here, the program instrumentation is useful in determining
which inputs to retain for inclusion in the test suite and further
fuzzing.

\subsection{Running Example}
\label{subsect:example}

Fig.~\ref{fig:runningExample} shows a simple, constructed program that
we will use throughout the paper to demonstrate the benefits of our
approach. Function \code{bar} takes as input three integers \code{a},
\code{b}, and \code{c} and returns an integer. There are five paths in
this function, all of which are feasible. Each path is denoted by a
unique return value. (The grey boxes are discussed below and should be
ignored for now.)

When running traditional greybox fuzzing (more specifically, AFL) on
function \code{bar}, only four out of five program paths are explored
within 12h. During this time, greybox fuzzing constructs a test suite
of four inputs, each of which explores a different path in \code{bar}.
The path with return value 2 remains unexplored even after the fuzzer
generates about 311M different inputs. All but four of these inputs
are discarded as they exercise a path in \code{bar} that has already
been covered by a previous test case.

\begin{figure}[t]
\begin{lstlisting}[style=clang, xleftmargin=1.5em]
int bar(a, b, c: int) {
  var d = b + c;
  °minimize(d < 1 ? 1 - d : 0);°  ¤\label{line:minimize1T}¤
  °minimize(d < 1 ? 0 : d);° ¤\label{line:minimize1F}¤
  if d < 1 { ¤\label{line:equalsPre1}¤
    °minimize(b < 3 ? 3 - b : 0);° ¤\label{line:minimize2T}¤
    °minimize(b < 3 ? 0 : b - 2);° ¤\label{line:minimize2F}¤
    if b < 3 { ¤\label{line:equalsPre2}¤
      return 1;
    }
    °minimize(a == 42 ? 1 : 0);° ¤\label{line:minimize3T}¤
    °minimize(a == 42 ? 0 : |a - 42|);° ¤\label{line:minimize3F}¤
    if a == 42 { ¤\label{line:equals}¤
      return 2; ¤\label{line:ret2}¤
    }
    return 3;
  } else {
    °minimize(c < 42 ? 42 - c : 0);° ¤\label{line:minimize4T}¤
    °minimize(c < 42 ? 0 : c - 41);° ¤\label{line:minimize4F}¤
    if c < 42 {
      return 4;
    }
    return 5;
  }
}
\end{lstlisting}
\vspace{-2em}
\caption{Running example.}
\label{fig:runningExample}
\vspace{-1.5em}
\end{figure}

The path with return value 2 is not covered because greybox fuzzers
randomly mutate program inputs (line~7 of
Alg.~\ref{alg:greyboxFuzzing}). More generally, it is challenging for
fuzzers to generate inputs that satisfy ``narrow checks'', that is,
checks that only become true for very few input values, like the check
on line~\ref{line:equals} of the running example. In this case, the
probability that the fuzzer will generate the value 42 for input
\code{a} is 1 out of $2^{32}$ for 32-bit integers. Even worse, to
cover the path with return value 2 (line~\ref{line:ret2}), the sum of
inputs \code{b} and \code{c} needs to be less than 1
(line~\ref{line:equalsPre1}) and \code{b} must be greater than or
equal to 3 (line~\ref{line:equalsPre2}).
As a result, several techniques have been proposed to guide greybox
fuzzing to satisfy such narrow checks, for instance, by selectively
applying whitebox fuzzing~\cite{StephensGrosen2016}.

In contrast, our greybox fuzzing approach is able to guide test case
generation toward optimal program executions \emph{without} any
program analysis or constraint solving. For instance, to ultimately
maximize path coverage, the grey boxes in
Fig.~\ref{fig:runningExample} define cost metrics that are minimized
when execution flips a branch condition. For this example, our
algorithm explores all five program paths within only 0.27s and after
generating only 372 different inputs.

\subsection{Approach}
\label{subsect:approach}

This effectiveness of our technique on the running example becomes
possible by making traditional greybox fuzzing a lighter shade of
grey. In other words, our fuzzing algorithm has access to more
information about the program under test than traditional greybox
fuzzing, which is enough to guide exploration toward specific
executions.

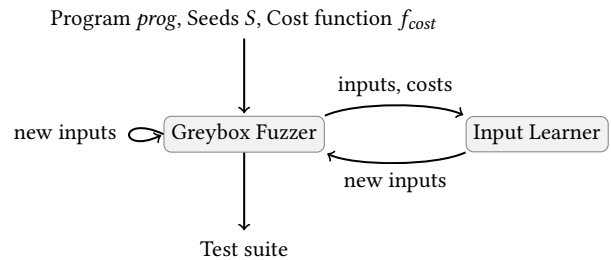
\begin{figure}[t]
\centering
\scalebox{0.95}{
  \begin{tikzpicture}[align=center, node distance=1.6cm]
    \node[draw=none] (P) at (0,0) {Program $\mathit{prog}$, Seeds $S$, Cost function $f_{\mathit{cost}}$};
    \node[draw, rounded corners=3, fill=black!5, draw=black!50, below of=P, yshift=0cm] (F) {Greybox Fuzzer};
    \node[draw, rounded corners=3, fill=black!5, draw=black!50, right of=F, xshift=2.5cm] (L) {Input Learner};
    \node[draw=none, below of=F, yshift=0cm] (T) {Test suite};
    \draw[thick, ->, shorten >=1pt] (P) -- (F);
    \draw[thick, ->, shorten >=1pt] (F) -- (T);
    \draw[thick, ->, shorten >=1pt, out=155, in=-155, looseness=10, loop, min distance=7mm] (F.west) to node[midway, anchor=center, fill=white, xshift=-0.9cm] {new inputs} (F.west);
    \draw[thick, ->, shorten >=1pt, out=25, in=155, looseness=0.6] (F.north east) to node[midway, anchor=center, fill=white, yshift=0.27cm] {inputs, costs} (L.north west);
    \draw[thick, ->, shorten >=1pt, out=-155, in=-25, looseness=0.6] (L.south west) to node[midway, anchor=center, fill=white, yshift=-0.27cm] {new inputs} (F.south east);
  \end{tikzpicture}
}
\caption{Overview of the workflow and tool architecture.}
\label{fig:workflow}
\vspace{-1em}
\end{figure}

The workflow and tool architecture of our greybox fuzzer is shown in
Fig.~\ref{fig:workflow}. The fuzzer takes as input the program under
test $\mathit{prog}$, a set of seeds $S$, and a partial function
$f_{\mathit{cost}}$ that maps program states to cost metrics.
When execution of $\mathit{prog}$ reaches a state $s$, the fuzzer
evaluates the cost metric $f_{\mathit{cost}}(s)$.
For example, the grey boxes in Fig.~\ref{fig:runningExample}
practically define a function $f_{\mathit{cost}}$ for the running
example. Each \code{minimize} statement specifies a cost metric at the
program state where it appears.
In other words, function $f_{\mathit{cost}}$ essentially constitutes
an instrumentation of the program under test, which is typically
derived automatically (see Sect.~\ref{sect:implementation}). In
comparison to traditional greybox fuzzing, this function is all the
additional information that our technique requires.

As shown in Fig.~\ref{fig:workflow}, the fuzzer passes explored inputs
and the corresponding costs, which are computed during execution of
$\mathit{prog}$ with these inputs, to a learning component. Based on
the given inputs and costs, this component learns new inputs that aim
to minimize the cost metrics of $f_{\mathit{cost}}$, and therefore,
lead to optimal executions.

As mentioned earlier, the cost metrics of
Fig.~\ref{fig:runningExample} define optimal executions as those that
flip a branch condition. Specifically, consider a program execution
along which variable \code{d} evaluates to 0. This execution takes the
then-branch of the first if-statement, and the cost metric defined by
the \code{minimize} statement on line~\ref{line:minimize1T} evaluates
to 1. This means that the distance of the current execution from an
execution that exercises the (implicit) else-branch of the
if-statement is 1. Now, consider a second program execution that also
takes the then-branch of the first if-statement (\code{d} evaluates to
-1). In this case, the cost metric on line~\ref{line:minimize1T}
evaluates to 2, which indicates a greater distance from an execution
that exercises the else-branch.

Based on this information, the learning component of
Fig.~\ref{fig:workflow} is able to learn new program inputs that force
execution of $\mathit{prog}$ to take the else-branch of the first
if-statement and minimize the cost metric on
line~\ref{line:minimize1T} (i.e., make the cost metric evaluate to
zero). For instance, the learned inputs could cause \code{d} to
evaluate to 7. Then, although the cost metric on
line~\ref{line:minimize1T} evaluates to zero, the cost metric on
line~\ref{line:minimize1F} evaluates to 7, which is the distance of
the current execution from an execution that exercises the then-branch
of the if-statement.

Similarly, the \code{minimize} statements on
lines~\ref{line:minimize2T}--\ref{line:minimize2F},
\ref{line:minimize3T}--\ref{line:minimize3F}, and
\ref{line:minimize4T}--\ref{line:minimize4F} of
Fig.~\ref{fig:runningExample} define cost metrics that are minimized
when a program execution flips a branch condition in a subsequent
if-statement. This instrumentation aims to maximize path coverage, and
for this reason, an execution can never minimize all cost metrics.
In fact, the fuzzer has achieved full path coverage when the generated
test cases cover all feasible combinations of branches in the program
under test; that is, when they minimize all possible combinations of
cost metrics.

As shown on the left of Fig.~\ref{fig:workflow}, the fuzzer may still
generate program inputs without the learning component, for instance,
when there are not enough executions from which to learn. In the above
example, the inputs for the first two program executions (where
\code{d} is 0 and -1) are generated by the fuzzer. Because the learner
can only approximate correlations between inputs and their
corresponding costs, it is possible that certain learned inputs do not
lead to optimal executions. In such cases, it is up to the fuzzer to
generate inputs that cover any remaining paths in the program under
test.

\section{Greybox Fuzzing with Learning}
\label{sect:technique}

In this section, we present the technical details of our approach.  In
particular, we describe our algorithm and explain how learning works.

\subsection{Algorithm}
\label{subsect:algorithm}

Alg.~\ref{alg:greyboxFuzzingWithLearning} shows how greybox fuzzing
with learning works. The grey boxes indicate the differences between
our technique and traditional greybox fuzzing
(Alg.~\ref{alg:greyboxFuzzing}).

In addition to the program under test $\mathit{prog}$ and a set of
seeds $S$, Alg.~\ref{alg:greyboxFuzzingWithLearning} takes as input a
partial function $f_{\mathit{cost}}$ that, as explained earlier, maps
program states to cost metrics. The fuzzer first runs the program with
the seeds, and during each program execution, it evaluates the cost
metric $f_{\mathit{cost}}(\mathit{s})$ for every encountered program
state $\mathit{s}$ in the domain of $f_{\mathit{cost}}$ (line~1). Like
in Alg.~\ref{alg:greyboxFuzzing}, each explored path in the program is
associated with a unique identifier $\mathit{PID}$. Note, however,
that the $\mathit{PIDs}$ data structure now maps a $\mathit{PID}$ both
to an input that exercises the corresponding path as well as to a cost
vector, which records all costs computed during execution of the
program with this input.
Next, an input is selected for mutation (line~3), and it is assigned
an energy value (line~5).

\begin{algorithm}[t]
  \caption{Greybox Fuzzing with Learning}
  \label{alg:greyboxFuzzingWithLearning}
  \hspace{-5em}\textbf{Input:} Program $\mathit{prog}$, Seeds $S$, {\btHL[fill=light-gray] Cost function $f_{\mathit{cost}}$}\\
  \begin{algorithmic}[1]
    \small
      \Let{$\mathit{PIDs}$}{\Fcall{RunSeeds}$(S, \mathit{prog},$ {\btHL[fill=light-gray] $f_{\mathit{cost}}$}$)$}
      \While{$\neg$\Fcall{Interrupted}()}
        \Let{$\mathit{input},$ {\btHL[fill=light-gray] $\mathit{cost}$}}{\Fcall{PickInput}$(\mathit{PIDs})$}
        \Let{$\mathit{energy}$}{0}
        \Let{$\mathit{maxEnergy}$}{\Fcall{AssignEnergy}$(\mathit{input})$}
        \LetHL{$\mathit{learnedInput}$}{\textbf{nil}}
        \While{$\mathit{energy} < \mathit{maxEnergy}$ {\btHL[fill=light-gray] $\vee \; \mathit{learnedInput} \neq$ \textbf{nil}}}
          \IfHL{$\mathit{learnedInput} \neq$ \textbf{nil}}
            \Indent
            \LetHL{$\mathit{input'}$}{$\mathit{learnedInput}$}
            \LetHL{$\mathit{learnedInput}$}{\textbf{nil}}
            \EndIndent
          \ElseHL
            \Indent
            \Let{$\mathit{input'}$}{\Fcall{FuzzInput}$(\mathit{input})$}
            \EndIndent
          \Let{$\mathit{PID'},$ {\btHL[fill=light-gray] $\mathit{cost'}$}}{\Fcall{Run}$(\mathit{input'}, \mathit{prog},$ {\btHL[fill=light-gray] $f_{\mathit{cost}}$} $)$}
          \If{\Fcall{IsNew}($\mathit{PID'}, \mathit{PIDs}$)}
            \Let{$\mathit{PIDs}$}{\Fcall{Add}$(\mathit{PID'}, \mathit{input'},$ {\btHL[fill=light-gray] $\mathit{cost'},$} $\mathit{PIDs})$}
          \EndIf
          \IfHL{$\mathit{energy} < \mathit{maxEnergy}$ }
            \Indent
            \LetHL{$\mathit{learnedInput}$}{\Fcall{Learn}$(\mathit{input},\mathit{cost},\mathit{input'},\mathit{cost'})$}
            \EndIndent
          \Let{$\mathit{energy}$}{$\mathit{energy} + 1$}
        \EndWhile
      \EndWhile
  \end{algorithmic}
  \hspace{-12em}\textbf{Output:} Test suite \textsc{Inputs}$(\mathit{PIDs})$\\
\end{algorithm}

The input is mutated (line~12), and the
program is run with the new input (line~13). For simplicity, we assume
that the new input differs from the original input (which was selected
for mutation on line~3) by the value of a single input parameter. As
usual, if the program follows a path that has not been previously
explored, the new input is added to the test suite (lines~14--15).

On line~17, the original input and the new input are passed to the
learning component along with their cost vectors. The learner inspects
$input$ and $input'$ to determine the input parameter by which they
differ. Based on the given cost vectors, it then learns a new value
for this input parameter such that one of the cost metrics is
minimized.
In case a new input is learned, the program is tested with this input,
otherwise the original input is mutated (lines~8--10). The former
happens even if the energy of the original input has run out (line~7)
in order to ensure that we do not waste learned inputs.

The above process is repeated until an exploration bound is reached
(line 2), and the fuzzer returns a test suite containing one test case
for each program path that has been explored.

\paragraph{\textbf{Running example}} In Tab.~\ref{tab:runningExample}, we run
our algorithm (Alg.~\ref{alg:greyboxFuzzingWithLearning}) on the
running example of Fig.~\ref{fig:runningExample} step by step. The
first column of the table shows an identifier for every generated test
case, and the second column shows the path that each test exercises
identified by the return value of the program. The highlighted boxes
in this column denote paths that are covered for the first time, which
means that the corresponding tests are added to the test suite
(lines~14--15 of Alg.~\ref{alg:greyboxFuzzingWithLearning}). The third
column shows the test identifier from which the value of variable
$\mathit{input}$ is selected (line~3 of
Alg.~\ref{alg:greyboxFuzzingWithLearning}). Note that, according to
the algorithm, $\mathit{input}$ is selected from tests in the test
suite. The fourth column shows a new input for the program under
test; this input is either a seed or the value of variable
$\mathit{input'}$ in the algorithm, which is obtained with learning
(line~9) or fuzzing (line~12). Each highlighted box in this column
denotes a learned value.  The fifth column shows the cost vector that
is computed when running the program with the new input of the fourth
column. Note that we show only non-zero costs and that the subscript
of each cost denotes the line number of the corresponding
\code{minimize} statement in Fig.~\ref{fig:runningExample}. The sixth
column shows which costs (if any) are used to learn a new input,
and the last column shows the current energy value of the algorithm's
$\mathit{input}$ (lines~4 and~18).
For simplicity, we consider $\mathit{maxEnergy}$ of
Alg.~\ref{alg:greyboxFuzzingWithLearning} (line~5) to always have the
value 2 in this example. Our implementation, however, is inspired by
an existing energy schedule~\cite{BoehmePham2016}.

We assume that the set of seeds $S$ contains only the random input
$(\text{\code{a}} = -1, \text{\code{b}} = 0, \text{\code{c}} = -5)$
(test \#1 in Tab.~\ref{tab:runningExample}). This input is then fuzzed
to produce $(\text{\code{a}} = -1, \text{\code{b}} = -3,
\text{\code{c}} = -5)$ (test \#2), that is, to produce a new value for
input parameter \code{b}. Our algorithm uses the costs computed with
metric $C_6$ to learn a new value for \code{b}. (We explain how new
values are learned in the next subsection.) As a result, test \#3
exercises a new path of the program (the one with return value
3). From the cost vectors of tests \#1 and \#3, only the costs
computed with metric $C_3$ may be used to learn another value for
\code{b}; costs $C_6$ and $C_7$ are already zero in one of the two
tests, while metric $C_{12}$ is not reached in test \#1. Even though
the energy of the original input (from test \#1) has run out, the
algorithm still runs the program with the input learned from the $C_3$
costs (line 7). This results in covering the path with return value 4.

Next, we select an input from tests \#1, \#3, or \#4 of the test
suite. Let's assume that the fuzzer picks the input from test \#3 and
mutates the value of input parameter \code{a}. Note that the cost
vectors of tests \#3 and \#5 differ only with respect to the $C_{12}$
costs, which are therefore used to learn a new input for \code{a}. The
new input exercises a new path of the program (the one with return
value 2). At this point, the cost vectors of tests \#3 and \#6 cannot
be used for learning because the costs are either the same ($C_3$ and
$C_7$) or they are already zero in one of the two tests ($C_{11}$ and
$C_{12}$). Since no input is learned and the energy of the original
input (from test \#3) has run out, our algorithm selects another input
from the test suite.

\begin{table}[t]
\centering
\scalebox{0.75}{
\begin{tabular}{c|c|c|ccc|c|c|c}
\multirow{2}{*}{\textsc{\textbf{Test}}} & \multirow{2}{*}{\textsc{\textbf{Path}}} & \textsc{\textbf{Input}} & \multicolumn{3}{c|}{\textsc{\textbf{New Input}}} & \multirow{2}{*}{\textsc{\textbf{Costs}}} & \textsc{\textbf{Learning}} & \multirow{2}{*}{\textsc{\textbf{Energy}}}\\
& & \textsc{\textbf{from Test}} & \texttt{a} & \texttt{b} & \texttt{c} & & \textsc{\textbf{Cost}} &\\
\hline
\multirow{2}{*}{1} & \cellcolor{light-gray} & \multirow{2}{*}{--} & \multirow{2}{*}{$-1$} & \multirow{2}{*}{0} & \multirow{2}{*}{$-5$} & $C_3 = 6$ & \multirow{2}{*}{--} & \multirow{2}{*}{--}\\
& \multirow{-2}{*}{\cellcolor{light-gray} 1} & & & & & $C_6 = 3$ & &\\
\hline
\multirow{2}{*}{2} & \multirow{2}{*}{1} & \multirow{2}{*}{1} & \multirow{2}{*}{$-1$} & \multirow{2}{*}{$-3$} & \multirow{2}{*}{$-5$} & $C_3 = 9$ & \multirow{2}{*}{$C_6$} & \multirow{2}{*}{0}\\
& & & & & & $C_6 = 6$ & &\\
\hline
\multirow{3}{*}{3} & \cellcolor{light-gray} & \multirow{3}{*}{1} & \multirow{3}{*}{$-1$} & \cellcolor{light-gray} & \multirow{3}{*}{$-5$} & $C_3 = 3$ & \multirow{3}{*}{$C_3$} & \multirow{3}{*}{1}\\
& \cellcolor{light-gray} & & & \cellcolor{light-gray} & & $C_7 = 1$ & &\\
& \multirow{-3}{*}{\cellcolor{light-gray} 3} & & & \multirow{-3}{*}{\cellcolor{light-gray} 3} & & $C_{12} = 43$ & &\\
\hline
\multirow{2}{*}{4} & \cellcolor{light-gray} & \multirow{2}{*}{1} & \multirow{2}{*}{$-1$} & \cellcolor{light-gray} & \multirow{2}{*}{$-5$} & $C_4 = 1$ & \multirow{2}{*}{--} & \multirow{2}{*}{2}\\
& \multirow{-2}{*}{\cellcolor{light-gray} 4} & & & \multirow{-2}{*}{\cellcolor{light-gray} 6} & & $C_{18} = 47$ & &\\
\hline
\multirow{3}{*}{5} & \multirow{3}{*}{3} & \multirow{3}{*}{3} & \multirow{3}{*}{7} & \multirow{3}{*}{3} & \multirow{3}{*}{$-5$} & $C_3 = 3$ & \multirow{3}{*}{$C_{12}$} & \multirow{3}{*}{0}\\
& & & & & & $C_7 = 1$ & &\\
& & & & & & $C_{12} = 35$ & &\\
\hline
\multirow{3}{*}{6} & \cellcolor{light-gray} & \multirow{3}{*}{3} & \cellcolor{light-gray} & \multirow{3}{*}{3} & \multirow{3}{*}{$-5$} & $C_3 = 3$ & \multirow{3}{*}{--} & \multirow{3}{*}{1}\\
& \cellcolor{light-gray} & & \cellcolor{light-gray} & & & $C_7 = 1$ & &\\
& \multirow{-3}{*}{\cellcolor{light-gray} 2} & & \multirow{-3}{*}{\cellcolor{light-gray} 42} & & & $C_{11} = 1$ & &\\
\hline
\multirow{2}{*}{7} & \multirow{2}{*}{4} & \multirow{2}{*}{4} & \multirow{2}{*}{$-1$} & \multirow{2}{*}{6} & \multirow{2}{*}{0} & $C_4 = 6$ & \multirow{2}{*}{$C_{18}$} & \multirow{2}{*}{0}\\
& & & & & & $C_{18} = 42$ & &\\
\hline
\multirow{2}{*}{8} & \cellcolor{light-gray} & \multirow{2}{*}{4} & \multirow{2}{*}{$-1$} & \multirow{2}{*}{6} & \cellcolor{light-gray} & $C_4 = 48$ & \multirow{2}{*}{--} & \multirow{2}{*}{1}\\
& \multirow{-2}{*}{\cellcolor{light-gray} 5} & & & & \multirow{-2}{*}{\cellcolor{light-gray} 42} & $C_{19} = 1$ & &\\
\end{tabular}}
\vspace{0.5em}
\caption{Running Alg.~\ref{alg:greyboxFuzzingWithLearning} on the
  example of Fig.~\ref{fig:runningExample}.}
\label{tab:runningExample}
\vspace{-2.5em}
\end{table}

This time, let's assume that the fuzzer picks the input from test \#4
and mutates the value of input parameter \code{c}. From the cost
vectors of tests \#4 and \#7, it randomly selects the $C_{18}$ costs
for learning a new value for \code{c}. The learned input exercises the
fifth path of the program, thus achieving full path coverage of
function \code{bar} by generating only 8 test cases.

Note that our algorithm makes several non-systematic choices, which
may be random or based on certain heuristics, such as when function
\textsc{PickInput} picks an input from the test suite, when
\textsc{FuzzInput} selects which input parameter to fuzz, or when
function \textsc{Learn} decides from which costs to learn. For
illustrating how the algorithm works, we made ``good'' choices such
that all program paths are exercised with a small number of test
cases. In practice, the fuzzer achieved full path coverage of function
\code{bar} with 372 test cases, instead of 8, as we discussed in
Sect.~\ref{subsect:example}.

\subsection{Learning}
\label{subsect:learning}

We now explain how the learning component of our approach works. Even
though this component is relatively simple, we found that it is very
effective in practice (see Sect.~\ref{sect:evaluation}). Note,
however, that our architecture (Fig.~\ref{fig:workflow}) is
configurable, and the learning component may be replaced with any
technique that learns new inputs for the program under test from
already explored inputs and the corresponding costs.

Our algorithm passes to the learning component the input vectors
$\mathit{input}$ and $\mathit{input'}$ and the corresponding cost
vectors $\mathit{cost}$ and $\mathit{cost'}$ (line~17 of
Alg.~\ref{alg:greyboxFuzzingWithLearning}). The input vectors differ
by the value of a single input parameter, say $i_0$ and $i_1$. Now,
let us assume that the learner selects a cost metric to minimize and
that the costs that have been evaluated using this metric appear as
$c_0$ and $c_1$ in the cost vectors. This means that cost $c_0$ is
associated with input parameter $i_0$, and $c_1$ with $i_1$.

As an example, let us consider tests \#3 and \#5 from
Tab.~\ref{tab:runningExample}. The input vectors that are passed to
the learner differ by the value of input parameter \code{a}, so $i_0 =
-1$ (value of \code{a} in test \#3) and $i_1 = 7$ (value of \code{a}
in test \#5). The learner chooses to learn from cost metric $C_{12}$
since the cost vectors of tests \#3 and \#5 differ only with respect
to this metric, so $c_0 = 43$ (value of $C_{12}$ in test \#3) and $c_1
= 35$ (value of $C_{12}$ in test \#5).

Using these two data points $(i_0, c_0)$ and $(i_1, c_1)$, we compute
the straight line $c(i) = m * i + k$ that connects them, where $m$ is
the slope of the line and $k$ is a constant. This line approximates
the relationship of the input parameter with the selected cost metric.
The line is computed (using basic algebra) as follows. The slope of
the line is:

\[
m = \frac{c_1 - c_0}{i_1 - i_0}
\]

\noindent
Once computed, we can solve for constant $k$.
Next, to compute the value that the input parameter should have in
order to minimize the cost metric, we solve the following equation for
$i$:

\[
m * i + k = 0 \Rightarrow i = - \frac{k}{m}
\]

From the data points $(-1, 43)$ and $(7, 35)$ defined by tests \#3 and
\#5, we compute the slope of the line to be $m = -1$ and the constant
to be $k = 42$. Therefore, the equation of the line connecting the two
data points is $c(i) = -i + 42$. Now, for the cost to be zero, the
value of parameter \code{a} must be:

\[
i = - \frac{42}{-1} \Rightarrow i = 42
\]

\noindent
Indeed, when \code{a} becomes 42 in test \#6, cost metric $C_{12}$ is
minimized.

However, as discussed earlier, the computed line \emph{approximates}
the relationship between an input parameter and the cost. This is
because costs are evaluated locally, at every branch condition,
without taking into account the structure of the program until that
point. As a result, our technique is very efficient in practice, but
there may be situations where the predicted value for an input
parameter fails to flip the target branch.
For example, consider the random seed input $(\text{\code{a}} = 0,
\text{\code{b}} = 0, \text{\code{c}} = 0)$ to the running example
($C_6 = 3$). Assume that input parameter \code{b} is then fuzzed to
produce $(\text{\code{a}} = 0, \text{\code{b}} = -1, \text{\code{c}} =
0)$ with $C_6 = 4$. According to the above, the line that connects the
two data points, involving parameter \code{b} and costs $C_6$, is
$c(i) = -i +3$, which means that for $C_6$ to be minimized, \code{b}
should have value 3. However, if \code{b} becomes 3, the branch
condition that we aimed to flip (on line~8 of
Fig.~\ref{fig:runningExample}) is not even reached; the condition on
line~5 is flipped instead.
We discuss the success rate of our technique in flipping the target
branches in Sect.~\ref{sect:evaluation}.

Our learning technique resembles linear regression with the difference
that we learn from only two data points. Our fuzzer could learn from
all previously explored inputs that differ by the value of the same
input parameter. However, storing all these inputs and then performing
look-up operations would most likely not scale for real programs,
where fuzzing can generate tens or hundreds of thousands of inputs in
only a few minutes (Sect.~\ref{sect:evaluation}).

\paragraph{\textbf{Selecting a cost metric for minimization}} Recall that,
in order to generate test \#8 in Tab.~\ref{tab:runningExample}, our
algorithm learns a new input from the cost vectors of tests \#4 and
\#7, comprising the $C_4$ and $C_{18}$ costs. In principle, the
algorithm could use the costs evaluated with either metric for
learning. In practice, however, it could use heuristics that, for
instance, pick from which costs to learn such that new branches are
flipped (like the branch on line~20 of Fig.~\ref{fig:runningExample},
which is flipped only by test \#8 after learning from the $C_{18}$
costs).

Alternatively, \code{minimize} statements could be augmented with a
second argument that specifies a \emph{weight} for the cost metric. A
comparison between such weights would then determine from which costs
to learn first, for instance, if $C_4$ had a greater weight than
$C_{18}$, the algorithm would prioritize the $C_4$ costs for
learning. Weights would also make it possible to impose a search
strategy on the greybox fuzzer, such as depth- or breadth-first
search.

\section{Implementation}
\label{sect:implementation}

In this section, we present the details of our implementation. We
describe the characteristics of the code that we target, namely of
Ethereum smart contracts, and discuss the specific cost metrics that
our fuzzer aims to minimize.

\subsection{Smart Contracts}
\label{subsect:contracts}

In recent years, there have emerged various general-use,
blockchain-based~\cite{BlockchainBlueprint,BlockchainTechnology,BlockchainRevolution},
distributed-computing platforms~\cite{BartolettiPompianu2017}, the
most popular of which is
\emph{Ethereum}~\cite{EthereumWhitePaper,Ethereum}. A key feature of
Ethereum is that it supports \emph{contract accounts} in addition to
user accounts, both of which publicly reside on the Ethereum
blockchain. Contract and user accounts store a balance and are owned
by a user. A contract account, however, is not directly managed by
users, but instead through code that is associated with it. This code
expresses contractual agreements between users, for example, to
implement and enforce an auction protocol. A contract account also has
persistent state (stored in a dictionary) that the code may access,
for instance, to store auction bids.
In order to interact with a contract account, users issue
\emph{transactions} that call functions of the account's code, for
instance, to place an auction bid. Executing a transaction requires
users to pay a fee, called \emph{gas}, which is roughly proportional
to how much code is run.

Contract accounts with their associated code and persistent state are
called \emph{smart contracts}. The code is written in a
Turing-complete bytecode, which is executed on the Ethereum Virtual
Machine (EVM)~\cite{EthereumYellowPaper}. Programmers, however, do not
typically write EVM code; they can instead program in several
high-level languages, like Solidity, Serpent, or Vyper, which compile
to EVM bytecode.

We have implemented a greybox fuzzer for smart contracts, which works
on the EVM-bytecode level and is being used commercially. We
implemented our learning technique on top of this fuzzer, and in our
experiments, we compare against the vanilla version (without
learning).  It is generally difficult to compare against other popular
off-the-shelf fuzzers, like AFL or LibFuzzer, since we target EVM
code. However, on the running example, AFL does not achieve full path
coverage after generating 311M inputs in 12h, whereas our vanilla
fuzzer exercises all paths after generating 37'354 inputs in 21.64s.

To test a smart contract, our tool generates, executes, and fuzzes
sequences of transactions, which call functions of the contract.  We
consider \emph{sequences} of transactions since each transaction may
have side effects on the contract's persistent state, which may affect
the execution of subsequent calls to the contract.

Note that function $f_{\mathit{cost}}$, introduced in
Sect.~\ref{subsect:approach}, constitutes a runtime instrumentation of
the contract under test. Although the \code{minimize} statements in
Fig.~\ref{fig:runningExample} give the impression that we instrument
the code at compile-time, we only use these statements for
illustration purposes. The design decision to use a runtime
instrumentation was taken because a compile-time instrumentation would
increase the gas usage of the contract and potentially lead to false
positives when detecting out-of-gas errors.

In this paper, we focus on two types of vulnerabilities in smart
contracts: (1)~crashes, which cause a transaction to be aborted and
waste assets paid as gas fees, and (2)~memory-access errors, which may
allow attackers to modify the persistent state of a contract. We
discuss these in more detail in the following sections.

\subsection{Cost Metrics}
\label{subsect:costMetrics}

We focus on two different kinds of cost metrics, (1)~ones that are
minimized when execution flips a branch condition, and (2)~ones that
are minimized when execution is able to modify arbitrary memory
locations.

\paragraph{\textbf{Branch conditions}} We have already discussed cost metrics
that are minimized when execution flips a branch condition in the
running example. Here, we describe how the cost metrics are derived
from the program under test.

For the comparison operators \code{==} ($\mathit{eq}$), \code{<}
($\mathit{lt}$), and \code{<=} ($\mathit{le}$), we define the
following cost functions:

\[
\begin{array}{lll}
  C_{\mathit{eq}}(l, r)  & = & \begin{cases} 1, & l = r\\ 0, & l \neq r \end{cases}\\
  C_{\mathit{\overline{eq}}}(l, r)  & = & \begin{cases} 0, & l = r\\ |l - r|, & l \neq r \end{cases}\\
  C_{\mathit{lt}}(l, r)  & = & \begin{cases} r - l, & l < r\\ 0, & l \geq r \end{cases}
\end{array}
\]
\[
\begin{array}{lll}
  C_{\mathit{\overline{lt}}}(l, r)  & = & \begin{cases} 0, & l < r\\ l - r + 1, & l \geq r \end{cases}\\
  C_{\mathit{le}}(l, r) & = & \begin{cases} r - l + 1, & l \leq r\\ 0, & l > r \end{cases}\\
  C_{\mathit{\overline{le}}}(l, r) & = & \begin{cases} 0, & l \leq r\\ l - r, & l > r \end{cases}
\end{array}
\]

\noindent
Function $C_{\mathit{eq}}$ from above is non-zero when a branch
condition $l$ \code{==} $r$ holds; it defines the cost metric for
making this condition false. On the other hand, function
$C_{\mathit{\overline{eq}}}$ defines the cost metric for making the
same branch condition true. The arguments $l$ and $r$ denote the left
and right operands of the operator. The notation is similar for all
other functions.

As an example, consider the branch condition on line~13 of
Fig.~\ref{fig:runningExample}, \code{a == 42}. According to function
$C_{\mathit{eq}}$, when this condition holds, the cost metric to make
it false is 1, as specified in the \code{minimize} statement on
line~11. This means that the distance from an execution where the
condition does not hold is 1, or in other words, if the value of
\code{a} changes by 1, the condition no longer holds. For an execution
where the condition on line~13 does not hold, function
$C_{\mathit{\overline{eq}}}$ defines the cost metric to make it true
as \code{|a - 42|} (see \code{minimize} statement on line~12).

Based on the above cost functions, our instrumentation evaluates two
cost metrics before every branch condition in the program under
test. The specific metrics that are evaluated depend on the
comparison operator used in the branch condition. The cost functions
for other comparison operators, namely \code{!=} ($\mathit{ne}$),
\code{>} ($\mathit{gt}$), and \code{>=} ($\mathit{ge}$), are easily
derived from the functions above. For instance, the cost functions for
the \code{!=} ($\mathit{ne}$) operator are defined as follows:

\vspace{-0.5em}
\[
\begin{array}{lll}
  C_{\mathit{ne}}(l, r)  & = & \begin{cases} |l - r|, & l \neq r\\ 0, & l = r \end{cases}\\
  C_{\mathit{\overline{ne}}}(l, r)  & = & \begin{cases} 0, & l \neq r\\ 1, & l = r \end{cases}\\
\end{array}
\]

\noindent
In other words:

\vspace{-0.5em}
\[
\begin{array}{lll}
C_{\mathit{ne}}(l, r) & \equiv & C_{\mathit{\overline{eq}}}(l, r)\\
C_{\mathit{\overline{ne}}}(l,r) & \equiv & C_{\mathit{eq}}(l, r)
\end{array}
\]

\noindent
Similarly:

\vspace{-0.5em}
\[
\begin{array}{lll}
C_{\mathit{gt}}(l, r) & \equiv & C_{\mathit{\overline{le}}}(l, r)\\
C_{\mathit{\overline{gt}}}(l, r) & \equiv & C_{\mathit{le}}(l, r)\\
C_{\mathit{ge}}(l, r) & \equiv & C_{\mathit{\overline{lt}}}(l, r)\\
C_{\mathit{\overline{ge}}}(l,r) & \equiv & C_{\mathit{lt}}(l, r)
\end{array}
\]

Note that our implementation works on the bytecode level, where
logical operators, such as \code{&&} and \code{||}, are typically
expressed as branch conditions. We, therefore, do not define cost
functions for such operators.

\paragraph{\textbf{Memory locations}} Recall that a smart
contract may store persistent state in a dictionary. If, however,
users or other contracts manage to access memory in this dictionary
(either accidentally or intentionally), they may endanger the
crypto-assets of the contract leading to critical security
vulnerabilities.

As an example, consider the smart contract (written in Solidity) shown
in Fig.~\ref{fig:memoryExample}. (The grey box should be ignored for
now.) It is a simplified version of code submitted to the Underhanded
Solidity Coding Contest (USCC) in 2017~\cite{USCC}. (In the next
section, we refer to the original code of this submission as
\textsc{USCC2}, with benchmark identifier~22.) The USCC is a contest
to write seemingly harmless Solidity code that, however, disguises
unexpected vulnerabilities.

\begin{figure}[t]
\begin{lstlisting}[style=sol, xleftmargin=1.5em]
contract Wallet {
  address private owner; ¤\label{line:field1}¤
  uint[] private bonusCodes; ¤\label{line:field2}¤

  constructor() public { ¤\label{line:constructor}¤
    owner = msg.sender;
    bonusCodes = new uint[](0);
  }

  function() public payable { } ¤\label{line:payable}¤

  function PushCode(uint c) public {
    bonusCodes.push(c);
  }

  function PopCode() public {
    require(0 <= bonusCodes.length); ¤\label{line:precondition1}¤
    bonusCodes.length--; ¤\label{line:overflow}¤
  }

  function SetCodeAt(uint i, uint c) public {
    require(i < bonusCodes.length);
    °minimize(|&(bonusCodes[i]) - 0xffcaffee|);° ¤\label{line:minimize}¤
    bonusCodes[i] = c; ¤\label{line:store}¤
  }

  function Destroy() public { ¤\label{line:destroy}¤
    require(msg.sender == owner); ¤\label{line:precondition2}¤
    selfdestruct(msg.sender); ¤\label{line:selfdestruct}¤
  }
}
\end{lstlisting}
\vspace{-1em}
\caption{Simplified example (written in Solidity) showing a
  vulnerability detected by greybox fuzzing with learning.}
\label{fig:memoryExample}
\vspace{-1em}
\end{figure}

The smart contract of Fig.~\ref{fig:memoryExample} implements a wallet
that has an owner and stores an array (with variable length) of bonus
codes (lines~\ref{line:field1}--\ref{line:field2}). The constructor
(on line~\ref{line:constructor}) initializes the owner to the address
of the caller and the bonus codes to an empty array. The empty
function on line~\ref{line:payable} ensures that assets can be payed
into the wallet. The following functions allow bonus codes to be
pushed, popped, or updated. The last function (on
line~\ref{line:destroy}) must be called only by the wallet owner
(line~\ref{line:precondition2}) and causes the wallet to self-destruct
(line~\ref{line:selfdestruct}), that is, to transfer all assets to the
owner and then destroy itself.

The vulnerability in this code is caused by the precondition on
line~\ref{line:precondition1}, which should require the length of the
array to be greater than zero (instead of greater than or equal to
zero) before popping an array element. Consequently, when the array is
empty, the statement on line~\ref{line:overflow} causes the (unsigned)
array length to underflow; this effectively disables the bound-checks
of the array, allowing elements to be stored anywhere in the
persistent storage of the contract. Therefore, by setting a bonus code
at a specific index in the array, an attacker could overwrite the
address of the wallet owner to their own address. Then, by destroying
the wallet, the attacker would transfer all assets to their
account. In a slightly more optimistic scenario, the wallet owner
could be accidentally set to an invalid address, in which case the
assets in the wallet would become inaccessible.

To detect such vulnerabilities, a greybox fuzzer can, for every
assignment to the persistent storage of a contract, pick an arbitrary
address and compare it to the target address of the assignment. When
these two addresses happen to be the same, it is very likely that the
assignment may also target other arbitrary addresses, perhaps as a
result of an exploit. A fuzzer (without learning) is only able to
detect these vulnerabilities by chance, and chances are extremely low
that the target address of an assignment matches an arbitrarily
selected address, especially given that these are 32 bytes long. In
fact, our greybox fuzzer without learning does not detect the
vulnerability in the code of Fig.~\ref{fig:memoryExample} within 12h.

To direct the fuzzer toward executions that could reveal such
vulnerabilities, we define the following cost function:

\vspace{-0.5em}
\[
  C_{\mathit{st}}(\mathit{lhsAddr}, \mathit{addr}) = |\mathit{lhsAddr} - \mathit{addr}|\\
\]

\noindent
In the above function, $\mathit{lhsAddr}$ denotes the address of the
left-hand side of an assignment to persistent storage (that is,
excluding assignments to local variables) and $\mathit{addr}$ an
arbitrary address. Function $C_{\mathit{st}}$ is non-zero when
$\mathit{lhsAddr}$ and $\mathit{addr}$ are different, and therefore,
optimal executions are those where the assignment writes to the
arbitrary address, potentially revealing a vulnerability.

Our instrumentation evaluates the corresponding cost metric before
every assignment to persistent storage in the program under test. An
example is shown on line~\ref{line:minimize} of
Fig.~\ref{fig:memoryExample}. (We use the \code{&} operator to denote
the address of \code{bonusCodes[i]}, and we do not show the
instrumentation at every assignment to avoid clutter.) Our fuzzer with
learning detects the vulnerability in the contract of
Fig.~\ref{fig:memoryExample} within a few seconds.


Detecting such vulnerabilities based on whether an assignment could
target an arbitrary address might generate false positives when the
arbitrary address is indeed an intended target of the assignment.
However, the probability of this occurring in practice is extremely
low (again due to the address length). In fact, we did not encounter
any false positives during our experiments.

\section{Experimental Evaluation}
\label{sect:evaluation}

In this section, we evaluate the effectiveness of greybox fuzzing with
learning on real-world smart contracts. First, we explain how we
selected our benchmarks (Sect.~\ref{sect:benchmark_selection}) and
describe the experimental setup we used (Sect.~\ref{sect:setup}). We
then show the results of comparing to traditional greybox fuzzing
(Sect.~\ref{sect:results}).

\subsection{Benchmark Selection}
\label{sect:benchmark_selection}

First, we collected all smart contracts from 16 GitHub
repositories. We selected the repositories based on two main criteria
to obtain a diverse set of contracts. On the one hand, we picked
popular projects in the Ethereum community (e.g., the Ethereum Name
Service domain auction, the Consensys wallet, and the MicroRaiden
payment service) and with high popularity on GitHub (3'675 stars in
total on 2018-05-02, median 106.5). The majority of the contracts in
these projects have been reviewed by independent auditors and are
deployed on the Ethereum blockchain, managing and transferring
significant amounts of crypto-assets on a daily basis. On the other
hand, we also selected repositories from a wide range of
application domains (e.g., auctions, token sales, payment networks,
and wallets) to cover various features of the Ethereum virtual machine
and of the Solidity programming language. We also included contracts
that had been hacked in the past (The DAO and the Parity wallet) and
five contracts (incl. the four top-ranked entries) from the repository
of the USCC to consider some malicious or buggy contracts.

\begin{table}[t!]
\centering
\scalebox{0.85}{
\begin{tabular}{r|l|r|r|l}
\multicolumn{1}{c|}{\textbf{BIDs}} & \multicolumn{1}{c|}{\textbf{Name}} & \multicolumn{1}{c|}{\textbf{Functions}} & \multicolumn{1}{c|}{\textbf{LOC}} & \multicolumn{1}{c}{\textbf{Description}}\\
\hline
1      & ENS   & 24  & 1205 & ENS domain name auction\\
2--3   & CMSW  & 49  & 503  & Consensys multisig wallet\\
4--5   & GMSW  & 49  & 704  & Gnosis multisig wallet\\
6      & BAT   & 23  & 191  & BAT token (advertising)\\
7      & CT    & 12  & 200  & Consensys token library\\
8      & ERCF  & 19  & 747  & ERC Fund (investment fund)\\
9      & FBT   & 34  & 385  & FirstBlood token (e-sports)\\
10--13 & HPN   & 173 & 3065 & Havven payment network\\
14     & MR    & 25  & 1053 & MicroRaiden payment service\\
15     & MT    & 38  & 437  & MOD token (supply-chain)\\
16     & PC    & 7   & 69   & Payment channel\\
17--18 & RNTS  & 49  & 749  & Request Network token sale\\
19     & DAO   & 23  & 783  & The DAO organization\\
20     & VT    & 18  & 242  & Valid token (personal data)\\
21     & USCC1 & 4   & 57   & USCC '17 entry\\
22     & USCC2 & 14  & 89   & USCC '17 (honorable mention)\\
23     & USCC3 & 21  & 535  & USCC '17 (3rd place)\\
24     & USCC4 & 7   & 164  & USCC '17 (1nd place)\\
25     & USCC5 & 10  & 188  & USCC '17 (2nd place)\\
26     & PW    & 19  & 549  & Parity multisig wallet\\
\end{tabular}
}
\vspace{0.5em}
\caption{Overview of benchmarks. The first column provides the
  benchmark IDs for each project in the second column. The third and
  fourth columns show the number of public functions and lines of code
  in the benchmarks.}
\label{tab:benchmarks}
\vspace{-2.5em}
\end{table}

Second, we identified one or more main contracts from each of the
repositories that would serve as contracts under test, resulting in a
total of 26 benchmarks. Note that many repositories contain several
contracts (incl. libraries) to implement a complex system, such as an
auction. Tab.~\ref{tab:benchmarks} gives an overview of all benchmarks
and the projects from which they originate. The first column lists the
benchmark IDs (BIDs) for each project, while the second column
provides the project name. The third and fourth columns show two
source code metrics to provide some indication of the complexity of
the contracts under test: the number of public functions in the
benchmarks and the lines of code (LOC). Note that all contracts were
written in Solidity. Finally, we provide a
short description of each project in the last
column. Appx.~\ref{sect:repos} provides more details about the
repositories and the tested changesets.

\subsection{Experimental Setup}
\label{sect:setup}

We ran our fuzzer both with and without learning on each of the 26
benchmarks. To compare the effectiveness of both approaches for
different usage scenarios, we selected three time limits: 5m (e.g.,
coffee break), 30m (e.g., lunch break), and 3h (e.g., over night). We
focus our comparison on the following two key metrics: path coverage
and number of detected bugs within a given time limit.

As explained in Sect.~\ref{subsect:contracts}, we focus on two types
of bugs here. On the one hand, we detect crashes due to assertion or
precondition violations; in addition to user-provided checks and
parameter validation, these include checked errors such as division by
zero or out-of-bounds array access inserted by the Solidity
compiler. In the best case scenario, these bugs cause a transaction to
be aborted and waste assets paid as gas fees. In the worst case
scenario, they prevent legitimate transactions from succeeding,
putting user assets at risk. For instance, a user may not be able to
claim ownership of an auctioned item due to an out-of-bounds error in
the code that iterates over an array of bidders to determine the
winner. On the other hand, we detect memory-access errors that may
allow an attacker to modify the persistent state of a contract (see
Fig.~\ref{fig:memoryExample} for an example). The fuzzers did not
report any spurious bugs.

We ran all experiments on an
Intel\textregistered~Xeon\textregistered~CPU~@~2.30GHz machine with 8
GB of memory running Ubuntu 17.10.

\subsection{Results}
\label{sect:results}

\begin{table}[t!]
\centering
\scalebox{0.8}{
\begin{tabular}{r|r|r|r|r|r|r|r|r|r}
\multicolumn{1}{c|}{\textbf{BID}} & \multicolumn{1}{c|}{$\text{\textbf{P}}$} & \multicolumn{1}{c|}{$\text{\textbf{P}}_{\text{\textbf{L}}}$} & \multicolumn{1}{c|}{$\text{\textbf{B}}$} & \multicolumn{1}{c|}{$\text{\textbf{B}}_{\text{\textbf{L}}}$} & \multicolumn{1}{c|}{$\text{\textbf{E}}$} & \multicolumn{1}{c|}{$\text{\textbf{E}}_{\text{\textbf{L}}}$} & \multicolumn{1}{c|}{$\text{\textbf{S}}_{\text{\textbf{L}}}$} & \multicolumn{1}{c|}{$\text{\textbf{F}}_{\text{\textbf{L}}}$} & \multicolumn{1}{c}{$\text{\textbf{R}}_{\text{\textbf{L}}}$}\\
\hline
1  & 94 & \textbf{127} & 0 & 0 & 46520 & 56615 & 4893 & 822 & 0.86\\
2  & 68 & \textbf{79} & 2 & 2 & 68790 & 50176 & 3349 & 577 & 0.85\\
3  & 49 & \textbf{91} & 2 & 2 & 73941 & 46660 & 2652 & 879 & 0.75\\
4  & 55 & \textbf{102} & 2 & 2 & 82998 & 36466 & 3430 & 1067 & 0.76\\
5  & 47 & \textbf{69} & 2 & 2 & 67424 & 40371 & 3305 & 608 & 0.84\\
6  & 51 & \textbf{52} & 0 & 0 & 107831 & 109290 & 7791 & 1257 & 0.86\\
7  & 28 & 28 & 0 & 0 & 39765 & 34324 & 937 & 569 & 0.62\\
8  & 39 & \textbf{74} & 0 & \textbf{1} & 123474 & 116255 & 3844 & 534 & 0.88\\
9  & 73 & \textbf{84} & 0 & 0 & 81399 & 83091 & 5074 & 720 & 0.88\\
10 & 55 & \textbf{162} & 0 & 0 & 28236 & 25236 & 3771 & 989 & 0.79\\
11 & 95 & \textbf{184} & 1 & 1 & 34253 & 34182 & 5157 & 1196 & 0.81\\
12 & 67 & \textbf{89} & 0 & 0 & 35915 & 37858 & 4713 & 663 & 0.88\\
13 & 41 & \textbf{77} & 4 & \textbf{5} & 33445 & 27155 & 3202 & 592 & 0.84\\
14 & 62 & \textbf{84} & 2 & 2 & 50908 & 51232 & 3509 & 1117 & 0.76\\
15 & 84 & \textbf{100} & 1 & 1 & 87209 & 82338 & 4539 & 1249 & 0.78\\
16 & \textbf{18} & 17 & 0 & 0 & 188235 & 186627 & 2750 & 327 & 0.89\\
17 & 63 & \textbf{79} & 0 & 0 & 46906 & 54769 & 5126 & 957 & 0.84\\
18 & 49 & \textbf{78} & \textbf{2} & 1 & 58092 & 48794 & 4089 & 1776 & 0.70\\
19 & 61 & \textbf{74} & 7 & 7 & 83636 & 89522 & 3695 & 597 & 0.86\\
20 & 51 & \textbf{68} & 0 & 0 & 119444 & 90801 & 4809 & 778 & 0.86\\
21 & 24 & \textbf{28} & 2 & 2 & 163260 & 199974 & 2146 & 251 & 0.90\\
22 & 39 & \textbf{40} & 2 & 2 & 173082 & 134151 & 4677 & 718 & 0.87\\
23 & 57 & \textbf{82} & 1 & \textbf{3} & 125518 & 82929 & 5927 & 2058 & 0.74\\
24 & 20 & \textbf{25} & 5 & \textbf{9} & 215256 & 194288 & 3011 & 377 & 0.89\\
25 & 22 & \textbf{34} & 0 & \textbf{1} & 158859 & 168585 & 4865 & 670 & 0.88\\
26 & 37 & \textbf{68} & 13 & \textbf{23} & 54264 & 36587 & 1960 & 2169 & 0.47
\end{tabular}
}
\vspace{1em}
\caption{Results of greybox fuzzing without and with learning for 26
  benchmarks (time limit of 5m).}
\label{tab:results300}
\vspace{-2em}
\end{table}

Tab.~\ref{tab:results300} shows the results of each approach with the
time limit of 5m. The first column identifies the benchmark, while the
second and third columns show the number of covered paths without
($\text{P}$) and with learning ($\text{P}_{\text{L}}$). We can observe that learning
increases coverage in 24 of 26 benchmarks, while only decreasing
coverage slightly for a single benchmark (benchmark 16). For most
benchmarks, the coverage increase is significant (up to 3X for
benchmark 10).

Similarly, learning allows the fuzzer to detect 19 bugs that are not
detected otherwise. The fourth and fifth columns of
Tab.~\ref{tab:results300} show the detected bugs without ($\text{B}$) and
with learning ($\text{B}_{\text{L}}$). In contrast, the fuzzer without learning only
finds a single bug that is not detected with learning (in benchmark
18).

To evaluate the potential overhead of learning, the sixth and seventh
columns also provide the total energy (i.e., number of generated
inputs) spent without ($\text{E}$) and with learning ($\text{E}_{\text{L}}$). There are only
9 benchmarks where learning is able to explore more inputs. This could
suggest that our approach generates fewer inputs within the time limit
due to the overhead of learning. However, note that, for most
benchmarks where fuzzing with learning achieves higher path coverage,
it does so with less energy. This shows that the quantity of inputs
alone is not a good indication of the effectiveness of these
techniques; better-quality inputs help achieve better results in the
same amount of time.
Moreover, because the two approaches explore different program inputs,
with varying running times, it is difficult to draw any meaningful
conclusion about their overhead based on the total energy.

\begin{table}[t!]
\centering
\scalebox{0.8}{
\begin{tabular}{r|r|r|r|r|r|r|r|r|r}
\multicolumn{1}{c|}{\textbf{BID}} & \multicolumn{1}{c|}{$\text{\textbf{P}}$} & \multicolumn{1}{c|}{$\text{\textbf{P}}_{\text{\textbf{L}}}$} & \multicolumn{1}{c|}{$\text{\textbf{B}}$} & \multicolumn{1}{c|}{$\text{\textbf{B}}_{\text{\textbf{L}}}$} & \multicolumn{1}{c|}{$\text{\textbf{E}}$} & \multicolumn{1}{c|}{$\text{\textbf{E}}_{\text{\textbf{L}}}$} & \multicolumn{1}{c|}{$\text{\textbf{S}}_{\text{\textbf{L}}}$} & \multicolumn{1}{c|}{$\text{\textbf{F}}_{\text{\textbf{L}}}$} & \multicolumn{1}{c}{$\text{\textbf{R}}_{\text{\textbf{L}}}$}\\
\hline
1  & 263 & \textbf{726} & 0 & 0 & 150704 & 75472 & 6848 & 2468 & 0.74\\
2  & 104 & \textbf{202} & 2 & 2 & 287151 & 277622 & 11468 & 3193 & 0.78\\
3  & 119 & \textbf{197} & 2 & 2 & 231886 & 294640 & 11279 & 4216 & 0.73\\
4  & 122 & \textbf{183} & 2 & 2 & 227011 & 213853 & 8957 & 5113 & 0.64\\
5  & 153 & 153 & 2 & 2 & 184248 & 153015 & 6712 & 3313 & 0.67\\
6  & 52 & \textbf{53} & 0 & 0 & 606372 & 595179 & 16473 & 3206 & 0.84\\
7  & 28 & 28 & 0 & 0 & 220830 & 209444 & 4445 & 1269 & 0.78\\
8  & 68 & \textbf{87} & \textbf{3} & 2 & 613358 & 539998 & 11793 & 2623 & 0.82\\
9  & 83 & \textbf{88} & 0 & 0 & 449961 & 491465 & 21086 & 4095 & 0.84\\
10 & 110 & \textbf{239} & 0 & \textbf{1} & 197927 & 158557 & 11608 & 6044 & 0.66\\
11 & 170 & \textbf{228} & 1 & 1 & 174498 & 221989 & 12973 & 4213 & 0.75\\
12 & 87 & \textbf{92} & 0 & 0 & 232018 & 212204 & 12030 & 4464 & 0.73\\
13 & 64 & \textbf{79} & 5 & 5 & 189378 & 170803 & 6095 & 1516 & 0.80\\
14 & 92 & \textbf{123} & 2 & 2 & 220199 & 236748 & 11476 & 5789 & 0.66\\
15 & 106 & \textbf{120} & 1 & \textbf{2} & 475638 & 558284 & 23598 & 6729 & 0.78\\
16 & 18 & 18 & 0 & 0 & 1045947 & 1019000 & 9161 & 1057 & 0.90\\
17 & 78 & \textbf{79} & 0 & 0 & 272274 & 377359 & 17257 & 4109 & 0.81\\
18 & 71 & \textbf{79} & 2 & 2 & 293611 & 308228 & 11851 & 6237 & 0.66\\
19 & 79 & \textbf{82} & 7 & 7 & 386309 & 453576 & 13844 & 2208 & 0.86\\
20 & 57 & \textbf{115} & 0 & \textbf{1} & 629243 & 411287 & 15762 & 3138 & 0.83\\
21 & 24 & \textbf{28} & 2 & 2 & 820862 & 1094561 & 6783 & 816 & 0.89\\
22 & 41 & \textbf{42} & 2 & \textbf{3} & 967129 & 843334 & 15063 & 3191 & 0.83\\
23 & 74 & \textbf{84} & 3 & \textbf{4} & 496172 & 439086 & 14425 & 5710 & 0.72\\
24 & 20 & \textbf{26} & 5 & \textbf{9} & 1213700 & 1045862 & 8035 & 1157 & 0.87\\
25 & 22 & \textbf{34} & 0 & \textbf{1} & 890448 & 944219 & 13502 & 2363 & 0.85\\
26 & 88 & \textbf{97} & 21 & \textbf{23} & 214565 & 110071 & 4253 & 13947 & 0.23
\end{tabular}
}
\vspace{1em}
\caption{Results of greybox fuzzing without and with learning (time limit of 30m).}
\label{tab:results1800}
\vspace{-2em}
\end{table}

The last three columns of the table provide more details on how often
learning succeeds in minimizing the intended cost. Columns $\text{S}_{\text{L}}$ and
$\text{F}_{\text{L}}$ show the number of successful and failed minimizations,
respectively. A failed minimization indicates that our approximation
of how inputs relate to costs is imprecise. Column $\text{R}_{\text{L}}$ provides the
success rate of learning. We can see that the success rate is 0.81
on average (between 0.47 and 0.9), which suggests that it is very
effective.

The lowest success rate is observed for benchmark 26. Manual
inspection of this benchmark revealed that the contract makes
extensive use of cryptographic hashes (even during input
sanitization), a known obstacle for most fuzzers. For instance,
exploring a branch with condition $\mathit{sha3}(x) = 42$ is very
challenging when $x$ is an input. Learning does not help in this case
since the cost of flipping the above branch cannot be approximated
using a polynomial, let alone a linear, function. Note, however, that
learning can be very effective for a branch with condition
$\mathit{sha3}(x) = y$, if both $x$ and $y$ are inputs. If we only
fuzz input $y$, the expression $\mathit{sha3(x)}$ essentially becomes
a constant and learning succeeds. Despite the relatively low success
rate for benchmark 26, learning is still much more effective in both
increasing coverage and detecting bugs.

By looking at the same results for the time limit of 30m
(Tab.~\ref{tab:results1800}), we can observe the same trends as for
the limit of 5m. First, learning clearly increases coverage in 23
benchmarks and achieves the same coverage for the others. Second,
learning helps detect more bugs in 8 benchmarks. In contrast, the
fuzzer without learning only finds one additional bug. The main
difference between Tabs.~\ref{tab:results300}
and~\ref{tab:results1800} is that we now observe a slightly lower
success rate of 0.73 on average. We suspect that this is due to the
fact that we are in a later fuzzing phase where deeper execution
traces are explored. This can decrease the probability of minimizing
an intended cost when inputs also affect earlier branches in the
program (see Sect.~\ref{subsect:learning} for an example). However, as
observed earlier, the effectiveness of learning does not seem to hinge
on very high success rates.

\begin{table}[t!]
\centering
\scalebox{0.8}{
\begin{tabular}{r|r|r|r|r|r|r|r|r|r}
\multicolumn{1}{c|}{\textbf{BID}} & \multicolumn{1}{c|}{$\text{\textbf{P}}$} & \multicolumn{1}{c|}{$\text{\textbf{P}}_{\text{\textbf{L}}}$} & \multicolumn{1}{c|}{$\text{\textbf{B}}$} & \multicolumn{1}{c|}{$\text{\textbf{B}}_{\text{\textbf{L}}}$} & \multicolumn{1}{c|}{$\text{\textbf{E}}$} & \multicolumn{1}{c|}{$\text{\textbf{E}}_{\text{\textbf{L}}}$} & \multicolumn{1}{c|}{$\text{\textbf{S}}_{\text{\textbf{L}}}$} & \multicolumn{1}{c|}{$\text{\textbf{F}}_{\text{\textbf{L}}}$} & \multicolumn{1}{c}{$\text{\textbf{R}}_{\text{\textbf{L}}}$}\\
\hline
1  & 1676 & \textbf{2333} & 0 & 0 & 294797 & 308163 & 18957 & 9016 & 0.68\\
2  & 254 & \textbf{452} & 2 & 2 & 1046123 & 1079040 & 30588 & 33326 & 0.48\\
3  & 306 & \textbf{422} & 2 & 2 & 865133 & 800498 & 23654 & 33444 & 0.41\\
4  & 268 & \textbf{498} & 2 & 2 & 833864 & 1229650 & 36689 & 33242 & 0.52\\
5  & 256 & \textbf{320} & 2 & 2 & 904658 & 891458 & 26586 & 33119 & 0.45\\
6  & 52 & 52 & 0 & 0 & 4097566 & 4011683 & 57385 & 14074 & 0.80\\
7  & 28 & 28 & 0 & 0 & 1391220 & 1330754 & 14011 & 6759 & 0.67\\
8  & 116 & \textbf{141} & 9 & \textbf{13} & 3507219 & 3229561 & 48816 & 18105 & 0.73\\
9  & 85 & \textbf{88} & 0 & 0 & 3021580 & 2799495 & 72604 & 15824 & 0.82\\
10 & 143 & \textbf{383} & 1 & \textbf{3} & 1083042 & 887115 & 35173 & 23128 & 0.60\\
11 & 230 & \textbf{362} & 1 & 1 & 1161930 & 1006539 & 56058 & 21007 & 0.73\\
12 & 93 & \textbf{95} & 0 & 0 & 1316578 & 1254882 & 40935 & 15647 & 0.72\\
13 & 77 & \textbf{86} & 7 & 7 & 1222064 & 1185722 & 25703 & 9146 & 0.74\\
14 & 178 & \textbf{200} & 2 & 2 & 924258 & 952175 & 35087 & 27784 & 0.56\\
15 & \textbf{247} & 244 & 3 & 3 & 2271856 & 1987588 & 66959 & 31137 & 0.68\\
16 & 18 & 18 & 0 & 0 & 5660256 & 5555393 & 42573 & 5048 & 0.89\\
17 & 122 & \textbf{132} & 0 & 0 & 1166845 & 1144578 & 38296 & 17339 & 0.69\\
18 & 74 & \textbf{90} & 3 & \textbf{4} & 2079860 & 2083114 & 45293 & 21738 & 0.68\\
19 & 94 & \textbf{104} & 7 & 7 & 1762548 & 2072935 & 49674 & 6948 & 0.88\\
20 & 149 & \textbf{194} & 1 & 1 & 1862167 & 1533119 & 45512 & 21766 & 0.68\\
21 & 24 & \textbf{28} & 2 & 2 & 4753217 & 6518330 & 27986 & 3770 & 0.88\\
22 & 41 & \textbf{42} & 2 & \textbf{3} & 5892858 & 5216357 & 63590 & 15279 & 0.81\\
23 & 86 & 86 & 4 & 4 & 2954445 & 2578311 & 45565 & 25320 & 0.64\\
24 & 20 & \textbf{34} & 5 & \textbf{9} & 7292754 & 6383068 & 34142 & 5656 & 0.86\\
25 & 22 & \textbf{34} & 0 & \textbf{1} & 5313742 & 5559319 & 59849 & 11199 & 0.84\\
26 & 118 & \textbf{130} & 23 & 23 & 540616 & 638028 & 13361 & 99340 & 0.12
\end{tabular}
}
\vspace{1em}
\caption{Results of greybox fuzzing without and with learning (time limit of 3h).}
\label{tab:results10800}
\vspace{-2em}
\end{table}

Finally, the results when running both fuzzers for 3h (Tab.~\ref{tab:results10800})
confirm the same trends observed for shorter running times. With learning, we achieve
higher coverage for 21 benchmarks and we find more bugs in 6 benchmarks. Benchmark 15 is
the only one where we achieve marginally higher coverage without learning.

\begin{figure}[t!]
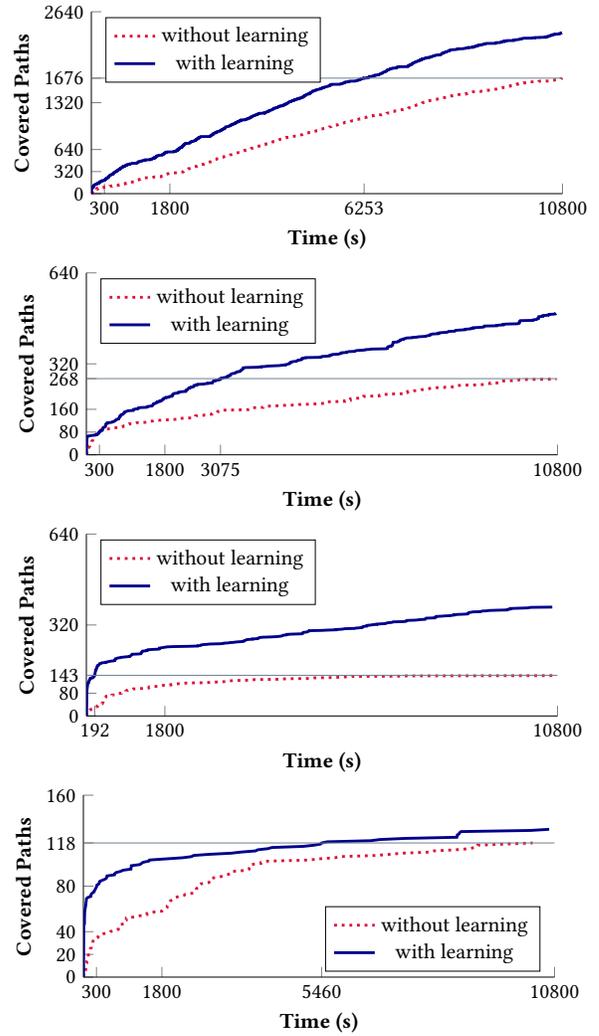

\centering
\scalebox{0.9}{

}
\vspace{-1em}
\caption{Coverage increase over time. These plots show how coverage
  increases over time for several benchmarks (1, 4, 10, and 26 from
  top to bottom) without (dotted line) and with learning (solid
  line). The light grey horizontal lines illustrate at which point in
  time one fuzzer surpasses the final coverage of the other.}
\label{fig:plots}
\vspace{-1em}
\end{figure}

\begin{table*}[t!]
\begin{minipage}[t][15cm]{0.47\textwidth}
\centering
\scalebox{0.8}{
\begin{tabular}{r|r|r|r|r}
\multicolumn{1}{c|}{\textbf{BID}} & \multicolumn{1}{c|}{\textbf{Bug}} & \multicolumn{1}{c|}{\textbf{$\text{\textbf{T}}$ (s)}} & \multicolumn{1}{c|}{\textbf{$\text{\textbf{T}}_{\text{\textbf{L}}}$ (s)}} & \multicolumn{1}{c}{\textbf{Bug Type}}\\
\hline
2  &        1 & 13.87 & \textbf{4.27} & assertion violation\\
2  &        2 & 349.21 & \textbf{6.23} & assertion violation\\
3  &        3 & 108.04 & \textbf{0.41} & assertion violation\\
3  &        4 & 80.21 & \textbf{8.05} & assertion violation\\
4  &        5 & 34.79 & \textbf{4.58} & assertion violation\\
4  &        6 & 30.79 & \textbf{5.40} & assertion violation\\
5  &        7 & 177.70 & \textbf{0.86} & assertion violation\\
\end{tabular}
}
\end{minipage}
\begin{minipage}[t][15cm]{0.47\textwidth}
\centering
\scalebox{0.8}{
\begin{tabular}{r|r|r|r|r}
\multicolumn{1}{c|}{\textbf{BID}} & \multicolumn{1}{c|}{\textbf{Bug}} & \multicolumn{1}{c|}{\textbf{$\text{\textbf{T}}$ (s)}} & \multicolumn{1}{c|}{\textbf{$\text{\textbf{T}}_{\text{\textbf{L}}}$ (s)}} & \multicolumn{1}{c}{\textbf{Bug Type}}\\
\hline
19 &       47 & 108.55 & \textbf{9.29} & assertion violation\\
19 &       48 & 28.85 & \textbf{11.68} & assertion violation\\
20 &       49 & \textbf{680.29} & 3848.78 & precondition violation\\
21 &       50 & 13.52 & \textbf{4.56} & precondition violation\\
21 &       51 & 39.23 & \textbf{8.51} & precondition violation\\
22 &       52 & \textbf{2.44} & 4.54 & assertion violation\\
22 &       53 & 76.48 & \textbf{5.05} & precondition violation\\
\end{tabular}
}
\end{minipage}
\vspace{-1em}
\caption{Time to bug (time limit of 3h). The third and fourth columns
  show the time in seconds until the bug was found without ($\text{\textbf{T}}$) and
  with learning ($\text{\textbf{T}}_{\text{\textbf{L}}}$). The last column shows the type of each bug.}
\label{tab:timeToCrash}
\vspace{-1.5em}
\end{table*}

Overall, our results demonstrate the effectiveness of our approach in
achieving higher coverage and detecting more bugs within the same
time. The former can be observed even more clearly when looking at how
converge increases over time. In Fig.~\ref{fig:plots}, we can see a
head-to-head comparison between the two approaches for a few
benchmarks. In all cases, our approach surpasses the final coverage of
greybox fuzzing without learning within a fraction of the allocated
time.

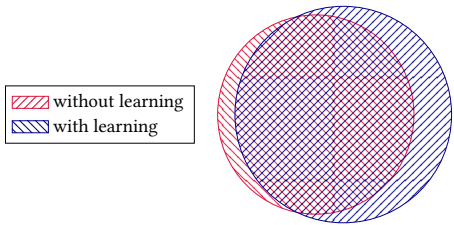
\begin{figure}[t]
\centering
  \scalebox{.8}{
  \begin{tikzpicture}[align=center]
  \begin{scope}[fill opacity=0.85,scale=0.6]
    \draw[Crimson, pattern color=Crimson, pattern=north west lines] (3.47910707406502,0.00) ellipse (2.71879001770779 and 2.76193581997921);
    \draw[DarkBlue, pattern color=DarkBlue, pattern=north east lines] (4.23968294364277,0.00) ellipse (3.00 and 3.00);
  \end{scope}
    \begin{customlegend}[legend entries={without learning,with learning},
        legend style={at={(-1.5,0.0)},anchor=center},
        legend cell align=left,]
    \begin{scope}[fill opacity=0.85]
      \addlegendimage{Crimson, pattern color=Crimson, pattern=north east lines,area legend}
      \addlegendimage{DarkBlue, pattern color=DarkBlue, pattern=north west lines,area legend}
    \end{scope}
    \end{customlegend}
  \end{tikzpicture}
  }
\vspace{-1em}
\caption{Venn diagram showing the average percentage of paths covered
  without (left circle) and with learning (right circle) for the time
  limit of 3h. On average, 7.1\% of paths are only covered without
  learning and 23.7\% with learning.}
\vspace{-1em}
\label{fig:venn}
\end{figure}

We also compare average percentages of covered paths in the Venn
diagram of Fig.~\ref{fig:venn}. In general, each approach covers paths
that are not covered by the other for some benchmarks. On average,
7.1\% of paths are only covered without learning and 23.7\% with
learning.

The effectiveness in finding bugs is even more apparent when comparing
the time until the same bug is detected (see
Tab.~\ref{tab:timeToCrash} for the time limit of 3h).
With learning, we find all bugs detected by fuzzing without learning
as well as 13 bugs more, for a total of 91 bugs. Moreover, our
technique finds 84 of these bugs within a significantly shorter time,
often orders of magnitude shorter. For instance, our technique finds
bug~7 roughly 200X faster and bug~76 600X faster.

\section{Threats to Validity}
\label{sect:threats}

We have identified the following threats to the validity of our
experiments.

\paragraph{\textbf{External validity}} With regard to external
validity~\cite{SiegmundSiegmund2015}, our experimental results may not
generalize to all smart contracts or to other types of programs. However,
we evaluated our technique on a diverse set of contracts from a wide
range of application domains. We, therefore, believe that our
benchmark selection significantly aids generalizability. In an effort
to further improve external validity, we also provide the versions of
all smart contracts we used in our experiments in
Appx.~\ref{sect:repos} so that others can test them. Independently,
our technique is not tailored to target smart-contract code
specifically; for this reason, we believe that our contributions
generalize to other languages.

Moreover, our comparison to greybox fuzzing without learning focuses
on one fuzzer, which we implemented. We discuss this below, as a way
of ensuring construct validity.

\paragraph{\textbf{Internal validity}} Another potential issue has to
do with the internal validity~\cite{SiegmundSiegmund2015} of our
experiments, that is, with whether systematic errors are introduced in
the experimental setup. A common threat to the internal validity of
experiments with fuzzing techniques is the selection of seeds. During
our experiments, when comparing greybox fuzzing with and without
learning, we always used the same seed inputs in order to avoid bias
in the exploration.

\paragraph{\textbf{Construct validity}} Construct validity ensures that
the experimental evaluation indeed measures what it claims, in our
case, the effect of our learning technique on greybox fuzzing. It is,
for instance, possible that engineering improvements to a tool provide
order-of-magnitude performance gains~\cite{RizziElbaum2016}. As an
example, consider the following two implementations: AFL does not
achieve full path coverage of the program of
Fig.~\ref{fig:runningExample} within 12h, whereas our greybox fuzzer
(without learning) does so in only 22s.

It is precisely for securing construct validity that we compare
against our own implementation of greybox fuzzing without
learning. The two fuzzers differ only in whether they use learning,
and as a result, we ensure that any improvements in the efficiency and
effectiveness of greybox fuzzing are exclusively the effect of our
technique. That is, no additional bias is introduced, for instance, by
different implementation details across fuzzers.

\section{Related Work}
\label{sect:relatedWork}

Our technique is the first to systematically learn new inputs for the
program under test with the goal of increasing the performance and
effectiveness of greybox fuzzing. The new inputs are learned from
previously explored inputs such that they guide fuzzing toward optimal
executions.

\paragraph{\textbf{Greybox fuzzing}} Greybox fuzzers~\cite{AFL,LibFuzzer}
rely on a lightweight program instrumentation to effectively discover
new program paths.
There are several techniques that aim to direct greybox fuzzing toward
certain parts of the search space, such as low-frequency
paths~\cite{BoehmePham2016}, vulnerable paths~\cite{RawatJain2017},
deep paths~\cite{SparksEmbleton2007}, or specific sets of program
locations~\cite{BoehmePham2017}. There are also techniques that intend
to boost fuzzing by smartly selecting and mutating
inputs~\cite{WooCha2013,RebertCha2014,ChaWoo2015}.

In general, greybox fuzzing with learning could be used in combination
with these techniques. In comparison, our approach learns concrete
input values from previous inputs, instead of performing arbitrary
input mutations. To achieve this, we rely on additional, but still
lightweight, instrumentation.

\paragraph{\textbf{Whitebox fuzzing}} Whitebox
fuzzers~\cite{GodefroidLevin2008,CadarDunbar2008}, like dynamic
symbolic execution tools~\cite{GodefroidKlarlund2005,CadarEngler2005},
use program analysis and constraint solving to understand the
structure of the program under test and, therefore, explore more
paths.
This approach has been implemented in many tools, such as
EXE~\cite{CadarGanesh2006}, jCUTE~\cite{SenAgha2006},
Pex~\cite{TillmanndeHalleux2008}, BitBlaze~\cite{SongBrumley2008},
Apollo~\cite{ArtziKiezun2010}, S2E~\cite{ChipounovKuznetsov2011}, and
Mayhem~\cite{ChaAvgerinos2012}, and comes in different flavors, such
as probabilistic symbolic execution~\cite{GeldenhuysDwyer2012} or
model-based whitebox fuzzing~\cite{PhamBoehme2016}. It has also been
tried in various application domains, such as testing software
patches~\cite{SantelicesChittimalli2008,BoehmeOliveira2013,MarinescuCadar2013},
complementing static analyzers~\cite{ChristakisMueller2016}, killing
higher-order mutants~\cite{HarmanJia2011}, and reconstructing field
failures~\cite{JinOrso2012,RoesslerZeller2013}.

As discussed earlier, our technique does not rely on any program
analysis or constraint solving and our instrumentation is more
lightweight, for instance, we do not keep track of a symbolic store
and path constraints.

\paragraph{\textbf{Hybrid fuzzing}} Hybrid fuzzers combine fuzzing with
other techniques to join their benefits and achieve better
results. For example, Dowser~\cite{HallerSlowinska2013} uses static
analysis to identify code regions with potential buffer
overflows. Similarly, BuzzFuzz~\cite{GaneshLeek2009} uses taint
tracking to discover which input bytes are processed by ``attack
points''. Hybrid Fuzz Testing~\cite{Pak2012} first runs symbolic
execution to find inputs that lead to ``frontier nodes'' and then
applies fuzzing on these inputs. On the other hand,
Driller~\cite{StephensGrosen2016} starts with fuzzing and uses
symbolic execution when it needs help in generating inputs that
satisfy complex checks.

In contrast, our approach extends greybox fuzzing without relying on
static analysis or whitebox fuzzing. It could, however, complement the
fuzzing component of these techniques to reduce the need for other
analyses.

\paragraph{\textbf{Optimization in testing}} Miller and
Spooner~\cite{MillerSpooner1976} were the first to use optimization
methods in generating test data, and in particular, floating-point
inputs. It was not until 1990 that these ideas were extended by Korel
for Pascal programs~\cite{Korel1990}. In recent years, such
optimization methods have been picked up again~\cite{McMinn2004},
enhanced, and implemented in various testing tools, such as
FloPSy~\cite{LakhotiaTillmann2010}, CORAL~\cite{SouzaBorges2011},
AUSTIN~\cite{LakhotiaHarman2013}, and CoverMe~\cite{FuSu2017}.

Most of these tools use fitness functions to determine how close the
current input is from a target. For instance, Korel uses fitness
functions that are similar to our cost metrics for flipping branch
conditions. The above tools search for input values that minimize
such functions. The search is typically iterative, for instance, by
using hill climbing or simulated
annealing~\cite{MetropolisRosenbluth1953,KirkpatrickGelatt1983}, and
it may not necessarily succeed in finding the minimum.  While our
technique also aims to minimize cost metrics, it does so in a single
shot. As a consequence, it may succeed faster, but it may also fail to
minimize the target metric altogether, in which case it falls back on
traditional greybox fuzzing.

\paragraph{\textbf{Program analysis for smart contracts}} The
program-analysis and verification community has already developed
several bug-finding techniques for smart contracts, including
debugging, static analysis, symbolic execution, and
verification~\cite{Remix,Securify,LuuChu2016,BhargavanDelignat-Lavaud2016,AtzeiBartoletti2017,ChenLi2017,SergeyHobor2017,ChatterjeeGoharshady2018,AmaniBegel2018,GrossmanAbraham2018,KalraGoel2018,NikolicKolluri2018}.

In contrast, our technique is the first to apply greybox fuzzing to
smart contracts.

\section{Conclusion}
\label{sect:conclusion}

In this paper, we have proposed a novel greybox fuzzing approach that
learns new inputs from existing program runs. These inputs are learned
such that they guide exploration toward optimal executions, as defined
by the given cost metrics. We demonstrate that learning significantly
increases the effectiveness of greybox fuzzing, both by achieving
higher path coverage and by detecting more bugs within the same time.
On a high level, our technique is an instance of a more general idea:
leveraging the large number of inputs explored by fuzzing to learn
more about the program under test.

In future work, we plan to investigate other instances of this general
idea by considering different cost metrics and by exploring
alternative learning techniques, such as decision-tree learning.


\newpage

\bibliography{tandem}

\newpage

\onecolumn

\appendix
\section{Smart Contract Repositories}
\label{sect:repos}

All analyzed smart contracts are available on GitHub. We provide\\
\noindent
changeset IDs and links to the repositories in
Tab.~\ref{tab:repos}.

\begin{table*}[b!]
\centering
\scalebox{0.9}{
\begin{tabular}{r|l|l|l}
\multicolumn{1}{c|}{\textbf{BIDs}} & \multicolumn{1}{c|}{\textbf{Name}} & \multicolumn{1}{c|}{\textbf{Changeset ID}} & \multicolumn{1}{c}{\textbf{Repository}}\\
\hline
1      & ENS   & 5108f51d656f201dc0054e55f5fd000d00ef9ef3 & \url{https://github.com/ethereum/ens}\\
2--3   & CMSW  & 2582787a14dd861b51df6f815fab122ff51fb574 & \url{https://github.com/ConsenSys/MultiSigWallet}\\
4--5   & GMSW  & 8ac8ba7effe6c3845719e480defb5f2ecafd2fd4 & \url{https://github.com/gnosis/MultiSigWallet}\\
6      & BAT   & 15bebdc0642dac614d56709477c7c31d5c993ae1 & \url{https://github.com/brave-intl/basic-attention-token-crowdsale}\\
7      & CT    & 1f62e1ba3bf32dc22fe2de94a9ee486d667edef2 & \url{https://github.com/ConsenSys/Tokens}\\
8      & ERCF  & c7d025220a1388326b926d8983e47184e249d979 & \url{https://github.com/ScJa/ercfund}\\
9      & FBT   & ae71053e0656b0ceba7e229e1d67c09f271191dc & \url{https://github.com/Firstbloodio/token}\\
10--13 & HPN   & 540006e0e2e5ef729482ad8bebcf7eafcd5198c2 & \url{https://github.com/Havven/havven}\\
14     & MR    & 527eb90c614ff4178b269d48ea063eb49ee0f254 & \url{https://github.com/raiden-network/microraiden}\\
15     & MT    & 7009cc95affa5a2a41a013b85903b14602c25b4f & \url{https://github.com/modum-io/tokenapp-smartcontract}\\
16     & PC    & 515c1b935ac43afc6bf54fcaff68cf8521595b0b & \url{https://github.com/mattdf/payment-channel}\\
17--18 & RNTS  & 6c39082eff65b2d3035a89a3f3dd94bde6cca60f & \url{https://github.com/RequestNetwork/RequestTokenSale}\\
19     & DAO   & f347c0e177edcfd99d64fe589d236754fa375658 & \url{https://github.com/slockit/DAO}\\
20     & VT    & 30ede971bb682f245e5be11f544e305ef033a765 & \url{https://github.com/valid-global/token}\\
21     & USCC1 & 3b26643a85d182a9b8f0b6fe8c1153f3bd510a96 & \url{https://github.com/Arachnid/uscc}\\
22     & USCC2 & 3b26643a85d182a9b8f0b6fe8c1153f3bd510a96 & \url{https://github.com/Arachnid/uscc}\\
23     & USCC3 & 3b26643a85d182a9b8f0b6fe8c1153f3bd510a96 & \url{https://github.com/Arachnid/uscc}\\
24     & USCC4 & 3b26643a85d182a9b8f0b6fe8c1153f3bd510a96 & \url{https://github.com/Arachnid/uscc}\\
25     & USCC5 & 3b26643a85d182a9b8f0b6fe8c1153f3bd510a96 & \url{https://github.com/Arachnid/uscc}\\
26     & PW    & 657da22245dcfe0fe1cccc58ee8cd86924d65cdd & \url{https://github.com/paritytech/contracts}
\end{tabular}
}
\vspace{1em}
\caption{Smart contract repositories.}
\label{tab:repos}
\end{table*}

\end{document}

%% file: solidity-highlighting.tex
\definecolor{verylightgray}{rgb}{.97,.97,.97}

\lstdefinelanguage{Solidity}{
	keywords=[1]{anonymous, assembly, assert, balance, break, call, callcode, case, catch, class, constant, constructor, continue, contract, debugger, default, delegatecall, delete, do, else, event, export, external, false, finally, for, function, gas, if, implements, import, in, indexed, instanceof, interface, internal, is, length, library, log0, log1, log2, log3, log4, memory, modifier, new, payable, pragma, private, protected, public, pure, push, require, return, returns, revert, selfdestruct, send, storage, struct, suicide, super, switch, then, this, throw, transfer, true, try, typeof, using, value, view, while, with, addmod, ecrecover, keccak256, mulmod, ripemd160, sha256, sha3}, 
	keywordstyle=[1]\color{DarkBlue}\bfseries,
	keywords=[2]{address, bool, byte, bytes, bytes1, bytes2, bytes3, bytes4, bytes5, bytes6, bytes7, bytes8, bytes9, bytes10, bytes11, bytes12, bytes13, bytes14, bytes15, bytes16, bytes17, bytes18, bytes19, bytes20, bytes21, bytes22, bytes23, bytes24, bytes25, bytes26, bytes27, bytes28, bytes29, bytes30, bytes31, bytes32, enum, int, int8, int16, int24, int32, int40, int48, int56, int64, int72, int80, int88, int96, int104, int112, int120, int128, int136, int144, int152, int160, int168, int176, int184, int192, int200, int208, int216, int224, int232, int240, int248, int256, mapping, string, uint, uint8, uint16, uint24, uint32, uint40, uint48, uint56, uint64, uint72, uint80, uint88, uint96, uint104, uint112, uint120, uint128, uint136, uint144, uint152, uint160, uint168, uint176, uint184, uint192, uint200, uint208, uint216, uint224, uint232, uint240, uint248, uint256, var, void, ether, finney, szabo, wei, days, hours, minutes, seconds, weeks, years},	
	keywordstyle=[2]\color{teal}\bfseries,
	keywords=[3]{block, blockhash, coinbase, difficulty, gaslimit, number, timestamp, msg, data, gas, sender, sig, value, now, tx, gasprice, origin},	
	keywordstyle=[3]\color{violet}\bfseries,
        keywords=[4]{minimize},
        keywordstyle=[4]\color{BrickRed}\bfseries,
	identifierstyle=\color{black},
	sensitive=false,
	comment=[l]{//},
	morecomment=[s]{/*}{*/},
	commentstyle=\color{gray}\ttfamily,
	stringstyle=\color{red}\ttfamily,
	morestring=[b]',
	morestring=[b]"
}